\documentstyle[aps]{revtex}
\begin{document}
\draft
\title{
Alternative Technique for  "Complex" Spectra Analysis }
\author{Pragya Shukla$^{*}$} 
\address{Department of Physics,   
Indian Institute of Technology, Kharagpur, India.}

\maketitle
\begin{abstract}
         The choice of a suitable random matrix model of a complex system is 
very sensitive to the nature of its complexity. The statistical 
spectral analysis of various complex systems requires, therefore, a thorough 
probing of a wide range of random matrix ensembles which is not an easy task. 
It is highly desirable, if possible,  to identify a common 
mathematical structure 
among all the ensembles and analyze it to gain information about the ensemble-
properties.
Our successful search in this direction leads to Calogero 
Hamiltonian,
a one-dimensional quantum 
Hamiltonian with inverse-square interaction, as the common base. 
This is because both, the  
  eigenvalues of the ensembles, and, a general state of Calogero Hamiltonian, 
  evolve in an analogous way for arbitrary initial conditions. The varying 
nature of the complexity is reflected in  different form of the evolution   
parameter in each case. A complete investigation of Calogero Hamiltonian 
can then help us in the spectral analysis of complex systems. 
\end{abstract}

	Recent statistical studies in various branches of theoretical 
physics, ranging from Calogero model of 1-d fermionic system \cite{cal}, 
random matrix (RM) model of disordered systems, 
matrix models of random surfaces to non-linear sigma 
model  
of quantum chaotic systems have revealed the presence of a common 
mathematical structure \cite{gmw,sla,ef}. 
The connecting-web of these various models with  
each other is well-described in \cite{sla}. 
However, so far, the connection of RM 
model with other models was established  only for standard Gaussian 
ensembles (SGE), that is, Gaussian 
ensembles invariant under unitary transformation. 
 This was achieved by 
showing that distribution of the eigenvalues of the ensemble is governed by a 
Fokker-Planck (F-P) equation \cite{fh,os}
similar to that of Dyson's "Brownian" motion 
model \cite{dy}.  
Through the reduction 
of F-P equation to the Schrodinger equation,  
the latter model is already known to be connected 
 to  Calogero Hamiltonian and  
thereby to various other models \cite{fh,os,bs}. In this paper, we explore   
RM models with non-invariant distributions, and, following the same route 
as in the case of SGE, connect them to Calogero Hamiltonian. This gives 
us a new technique to analyze the spectral behavior of the quantum 
operators of complex systems.

	The connection between Complex systems and Calogero 
Hamiltonian seems to 
be wide-ranging. 
	The eigenvalue dynamics  of Hermitian operators, for example, 
 Hamiltonians
of complex quantum systems e.g. chaotic systems, disordered systems 
seems to have an intimate connection with the particle-dynamics of 
Calogero-Moser (CM) Hamiltonian. The latter describes the 
 dynamics of particles interacting via pairwise 
  inverse square interaction and confined to move along a real line 
\cite{cal},  

\begin{eqnarray}
\hat H = 
- \sum_i{\partial^2 \over \partial\mu_i^2}
+ {1\over 4}\sum_{i<j}{\beta (\beta-2)\over (\mu_i-\mu_j)^2}-
\sum_i V(\mu_i) 
\end{eqnarray}

here $\mu_i$ is the position of the $i^{\rm th}$ particle and $V(\mu_i)$ 
is the confining potential. 
Similarly the level-dynamics of the unitary operators  
e.g. time-evolution operator is connected to 
Calogero-Sutherland (CS) Hamiltonian \cite{sut}:

\begin{eqnarray}
\hat H &=& 
-\sum_i{\partial^2 \over \partial\mu_i^2}
+{\beta(\beta-2)\over 16}
\sum_{i\not=j}{\rm cosec}^2 \left({\mu_i-\mu_j\over 2}\right)
-{\beta^2\over 48} N(N^2-1)
\end{eqnarray}

where particles are confined to move in a circle thus mimicking the 
similar confinement of eigenvalues due to unitary nature of the operator.  
The morphological transition caused by the interacting steps on a miscut 
crystal surface can also be modeled by the CS Hamiltonian \cite{las}. 
Here the  
 complexity is thermodynamic in nature.  
It is already well-known that the  parametric dispersion of 
the eigenvalues of the quantum system, with non-integrable classical limit,   
is described by a set of equations similar to the equations of motions of 
 particles, in time, of classical Calogero Hamiltonian \cite{pec}. This 
analogy extends also to the statistical properties in the two cases. 
 The parametric-evolution of the distribution 
$P_{N\beta}(\epsilon_1,..,\epsilon_N ;\tau)$ of the eigenvalues $\epsilon_i$  
of a  Hamiltonian $H=H_0 + \tau H_1$ (of size $N$),  
 with perturbation $H_1$ taken from a SGE   
 corresponds to the time-evolution of the distribution   
$P_{N\beta}(r_1,..,r_N ; t)$ of positions $r_i$'s of the particles and both the 
static as well as  dynamical correlators of the eigenvalues turn out to 
be similar to those of the particles in CM Hamiltonian (with 
$V(\mu) \propto \mu^2$ in eq.(1))\cite{fh,os,bs}.  
Here $\beta$ refers to the 
generic symmetry-class of the 
complex systems and therefore the transformation properties of 
associated RM models (known as GOE, GUE and GSE for $\beta=1,2$ and $4$) 
\cite{meta,fh}.
In limit $\tau \rightarrow \infty$, 
the eigenvalues attain an equilibrium distribution, known as Wigner-Dyson,  
 which  
 coincides with the probability 
distribution of $N$-particle coordinates $P_{N\beta}(r; t\rightarrow \infty)$ 
of the ground state 
of the CM Hamiltonian \cite{sla,bs}. 
Similar analogies can also be made between  evolution of the eigenvalues 
of unitary operators $U=U_0 {\rm e}^{i\tau M}$, with $M$ taken from SGE, 
 and CS Hamiltonian \cite{ap}. 
  This is  equivalent to say that the 
stationary and non-stationary states 
of CSM Hamiltonian  correspond to the eigenvalue distribution of  
the systems subjected to random perturbations,  
strong ($\tau \rightarrow \infty$) and weak (finite $\tau$) respectively,  
and thereby to equilibrium and non-equilibrium distribution of SGE.  
 In this paper, we indicate towards a
novel connection between the CM and RM model: a non-stationary state 
(finite $t$) of 
CM Hamiltonian can also be mapped to  
 the eigenvalue distribution of a 
generalized Gaussian ensemble (GGE); the correspondence is established 
by identifying a parameter $Y$ for GGE, equivalent to time $t$ for 
CM Hamiltonian.  
This mapping  
can then be used to obtain the information about various spectral 
properties of GGEs.    

	In recent past, RM ensembles have been quite often used to model 
the physical systems with complicated interactions \cite{gmw,meta}. 
The logic which could be  given in 
support of the model is that the missing information about the detailed 
nature of the interactions can be mimicked by randomizing the associated 
generators of motion, that is, by taking their matrix-representations as 
random matrices. 
However as the 
 specific details of the complexity of an operator should be reflected 
 in the associated RM model, the distribution of the matrix elements can 
 be of various types. For example, for a Hamiltonian with 
chaotic classical limit 
(least predictability of the long-term dynamics),  
 the distribution can be chosen as Gaussian (the 
least information 
 ensemble), with distribution parameters to be determined by the 
associated quantum dynamics. The corresponding RM model will thus belong 
to a generalized Gaussian ensemble  
 with matrix elements distribution given by 
$P(H) \propto {\rm e}^{-f_1(H).f_2(H)}$ (with $f_1$ and $f_2$ arbitrary 
functions and $H$ as a typical matrix).
The SGEs, with matrix elements distribution given by 
$P(H) \propto {\rm e}^{-Tr H^2}$  are special cases of GGEs
and many of their properties are already 
known. The various features of GGEs have,  however,  
remained unknown so far. The 
purpose of this paper is to suggest a technique to fill in this gap in 
our information. As for SGE, the nature of matrix elements in GGE too 
depends on the exact symmetry conditions of the Hamiltonian and is again 
indicated by parameter $\beta$, with $\beta=1, 2, 4$ for a generic matrix 
element to be real, complex or quaternion \cite{meta}. 
Here we discuss, in detail,  the properties of 
GG ensemble of complex hermitian matrices ($\beta=2$); 
the  GG ensemble of real-symmetric matrices  ($\beta=1$)has been discussed 
elsewhere \cite{ps1}.  We also probe briefly the non-Gaussian ensembles 
which can serve as good models for complex systems with various conditions 
on the associated quantum dynamics.

We proceed as follows. Our technique is based on  
 the statistical evolution of 
the eigenvalues of a GG ensemble with respect to a change in their distribution 
parameters. This requires a prior information  about the effect of a  
small change  in the matrix element on eigenvalues and eigenvectors; 
 the  related study  is  
given in section I.A. These results are then used to obtain, as described 
in section I.B, 
the distribution of eigenvalues 
$P(\mu,Y)$ of a matrix $H$ taken from a Gaussian ensemble, non-invariant 
under unitary transformation. 
The evolution of the eigenvalues is governed by a 
 partial differential equation which, after certain parametric redefinitions, 
turns out to be formally the 
same as the F-P equation for the Brownian motion of particles in  
Wigner-Dyson (WD) gas \cite{meta}.
The section II contains the details of the reduction of the F-P equation to 
the Schrodinger equation of CM Hamiltonian and a mapping of their respective 
correlators. 
The section III deals with the application of our technique to 
some important 
physical processes e.g. localization and  a brief 
discussion of our technique applied to a few other important matrix 
ensembles is given in section IV..  
We conclude in section VI which is 
followed by  the  appendices containing the proofs of some of the results 
given in the main text of the paper. 

\section{Eigenvalue Distribution of Generalized Gaussian Ensembles}

\subsection{The Change of Eigenvalues and Eigenfunctions} 

	The eigenvalue equation of a complex Hermitian matrix $H$ is given by 
$H U = U \Lambda$ with $\Lambda$ as the matrix of eigenvalues $\lambda_n$ 
and $U$ as 
the eigenvector 
matrix, unitary in nature. As obvious, a slight variation of the 
matrix elements 
of $H$ will, in general, lead to variation of both the eigenvalues as well as 
the eigenvectors and  associated rates of change can be obtained as follows; 

As $\lambda_n = \sum_{i,j} U_{ni} H_{ij} U_{nj}^*$,  
 the rate of change of $\lambda_n$ 
with respect to $H_{kl;s}$ (with $s$ referring to real, $s=1$, and imaginary, 
$s=2$, parts of $H_{kl}$) can be given  

\begin{eqnarray}
{\partial \lambda_n \over \partial H_{kl;s}}= 
   {i^{s-1}\over g_{kl}} [U_{ln} U_{kn}^{*} - (-1)^s U_{ln}^{*} U_{kn} ].  
\end{eqnarray}

where $g_{kl}=1+\delta_{kl}$. This
 can further be used to obtain the following relations (Appendix A)

\begin{eqnarray}
\sum_{k\le l}
 \sum_{s=1}^2 {\partial \lambda_n \over\partial H_{kl;s}} H_{kl;s}  
&=& \sum_{k, l}  
 H_{kl} U_{ln} U_{kn}^{*} = \lambda_n 
\end{eqnarray}
and 
\begin{eqnarray}
\sum_{k\le l} g_{kl}
 \sum_{s=1}^2 {\partial \lambda_n \over\partial H_{kl;s}}   
  {\partial \lambda_m \over\partial H_{kl;s}}  
 = 2\delta_{mn}
\end{eqnarray}

For our analysis later, we also require the information about 
 the second order  
change of an eigenvalue with respect to a matrix element and, therefore,  
 the rate of change of one of the eigenvector components with respect 
to $H_{kl}$. This is given as follows (Appendix B),   

\begin{eqnarray}
{\partial U_{pn} \over\partial H_{kl;s}}=
{i^{s-1} \over g_{kl}} \sum_{m\not=n}
{1\over {\lambda_n -\lambda_m}}
U_{pm}(U^{*}_{km}U_{ln} + (-1)^{s+1} U^{*}_{lm} U_{kn})
\end{eqnarray}

and now by using eqs.(3,6), One can show that (Appendix C)

\begin{eqnarray}
\sum_{k\le l} g_{kl} 
 \sum_{s=1}^2 {\partial^2 \lambda_n \over\partial H_{kl;s}^2}   
&=& 4 \sum_{m} {1 \over \lambda_n - \lambda_m}
\end{eqnarray}

	For the real-symmetric case, the corresponding relations 
can be obtained by using $U^+=U^T$ (as eigenvector matrix is 
now orthogonal) in eqs.(3-7) and taking $H_{ij;2}=0$ for all 
values of ${i,j}$ (see \cite{bs}). 

..
..

\subsection{The Evolution Equation For  
 the Eigenvalues}

Let us consider an ensemble of complex Hermitian matrices $H$, 
 with matrix elements $H_{kl}=H_{kl;1}+i H_{kl;2} (1-\delta_{kl})$ 
distributed as Gaussians with 
arbitrary variances and mean-values; the variances of  real and imaginary 
parts of a single matrix element  also need not be same.  
Thus we choose the distribution 
$\rho (H)$ of matrix $H$ to be following: 
\begin{eqnarray}
 \rho (H,y,b)=C{\rm exp}({ 
- \sum_{s=1}^2 \sum_{k\le l} \alpha_{kl;s}
 (H_{kl;s}- b_{kl;s})^2 })   
\end{eqnarray}
 with $C=\prod_{k\le l} \prod_{s=1}^2 \sqrt{\alpha_{kl;s}\over \pi}$ 
as the normalization constant,   
 $y$  as the set  
 of the coefficients  
$y_{kl;s}=\alpha_{kl;s} g_{kl}={g_{kl}\over 2 <H^2_{kl;s}>}$ 
 and $b$ as the set of all $b_{kl;s}$. 
Note that such a choice leads to a 
non-random complex Hamiltonian ($H_{kl}= b_{kl;1} + i b_{kl;2}$) 
in limit $\alpha_{kl;1}, \alpha_{kl;2} \rightarrow \infty$ and 
therefore can model various 
real physical situations such as switching of disorder in a non-random
Hamiltonian e.g. metal-insulator transitions.

Let $P(\mu,y,b)$ be the probability of finding eigenvalues 
$\lambda_i $  of $H$ between $\mu_i$ and $\mu_i+{\rm d}\mu_i$ at 
a given $y$ and $b$,
\begin{eqnarray}
P(\mu,y,b)= \int
\prod_{i=1}^{N}\delta(\mu_i-\lambda_i) \rho (H,y,b){\rm d}H 
\end{eqnarray}
 As the $\alpha$-dependence of $P$ 
in eq.(9) enters only through $\rho(H)$ and 
 ${\partial \rho \over\partial \alpha_{kl;s}} 
=\left[ (2 \alpha_{kl;s})^{-1} -(H_{kl;s}-b_{kl;s})^2\right]\rho
=(2 \alpha_{kl;s})^{-1}\left[ \rho + (H_{kl;s} - b_{kl;s}) 
{\partial \rho \over\partial H_{kl;s}} \right]$ with
${\partial \rho \over\partial H_{kl;s}}  
=- {\partial \rho \over\partial b_{kl;s}} $, 
a derivative of $P$ with respect to $\alpha_{kl;s}$  can be 
written as follows   

\begin{eqnarray}
{\partial P\over\partial \alpha_{kl;s}}  
& = & {P\over 2\alpha_{kl;s}}  
+ {1\over 2\alpha_{kl;s}}
\int \prod_{i=1}^N \delta(\mu_i-\lambda_i) 
H_{kl;s}{\partial\rho \over \partial H_{kl;s}} 
{\rm d}H   
+ {1\over 2\alpha_{kl;s}}
\int \prod_{i=1}^N \delta(\mu_i-\lambda_i) 
b_{kl;s}{\partial \rho \over \partial b_{kl;s}}
{\rm d}H 
\end{eqnarray}

The second integral in eq.(10) is equal to 
 $b_{kl;s}{\partial P \over \partial b_{kl;s}}$.   
The first integral can also be simplified by using 
integration by parts      
followed by a use of the equality
${\partial\prod_{i=1}^N\delta(\mu_i-\lambda_i)\over \partial H_{kl;s}}= 
-\sum_{n=1}^N {\partial 
\prod_{i=1}^N\delta(\mu_i-\lambda_i) 
\over \partial \mu_n}{\partial\lambda_n \over \partial H_{kl;s}}$: 

\begin{eqnarray}
\int \prod_{i=1}^N \delta(\mu_i-\lambda_i) 
H_{kl;s}{\partial\rho \over \partial H_{kl;s}} 
{\rm d}H &=&    
-\int {\partial \prod_{i=1}^N \delta(\mu_i-\lambda_i)\over \partial H_{kl;s}} 
\; H_{kl;s}\; \rho \;{\rm d}H 
-\int \prod_{i=1}^N \delta(\mu_i-\lambda_i) \rho\; {\rm d}H
= I_{kl;s} - P 
\end{eqnarray}
where 
\begin{eqnarray}
 I_{kl;s}&=&\sum_{n=1}^N {\partial \over \partial \mu_n} 
\int \prod_{i=1}^N 
\delta(\mu_i-\lambda_i) 
{\partial \lambda_n \over \partial H_{kl;s}} 
 H_{kl;s}\; \rho \;{\rm d}H
\end{eqnarray}
 
Substitution of eq.(11) in eq.(10) then gives 
\begin{eqnarray}
2 \alpha_{kl;s} {\partial P\over\partial \alpha_{kl;s}} = 
 I_{kl} + 
b_{kl;s}  
 {\partial P\over\partial b_{kl;s}}  
\end{eqnarray}

Our aim is to find a function $Y$ of the coefficients $\alpha_{kl;s}$'s 
and $b_{kl;s}$'s such 
that the evolution of $P(\mu,Y)$ in terms of $Y$ satisfies a F-P 
equation similar to that of Dyson's Brownian motion model (Wigner-Dyson gas) 
\cite{dy,meta}. 
For this purpose, we consider the sum  
2 $\sum_{k \le l}\left(\gamma - 
g_{kl}\alpha_{kl;s}\right)\alpha_{kl;s} 
{\partial P\over\partial \alpha_{kl;s}}$ where 
$\gamma$ is an arbitrary parameter 
and thereby obtain following relation

\begin{eqnarray}
\sum_{s=1}^2 \sum_{k \le l}\left(\gamma - y_{kl;s}\right)\left[2 y_{kl;s}
{\partial P\over\partial y_{kl;s}}  
- b_{kl;s} {\partial P\over\partial b_{kl;s}} \right]  
=\sum_{s=1}^2 \left[\gamma \sum_{k\le l}  I_{kl;s} -
\sum_{k\le l} y_{kl;s} I_{kl;s}\right]
\end{eqnarray}

As shown in Appendix D, the first term on the right hand side of eq.(14) 
can further be simplified,   

\begin{eqnarray}
 \sum_{s=1}^2 \sum_{k\le l}  I_{kl;s} =
\sum_{n}{\partial \over \partial\mu_n}\left(\mu_n P\right)
\end{eqnarray}


The second term can similarly be rewritten as follows (Appendix E):
\begin{eqnarray}
\sum_s \sum_{k\le l} y_{kl;s}  I_{kl;s}
= -\sum_n {\partial \over \partial\mu_n}
\left[ {\partial \over \partial\mu_n} +
\sum_{m\not=n}{\beta \over {\mu_m-\mu_n}}\right] P - \sum_{k\le l}  
y_{kl;s} b_{kl;s} {\partial P \over \partial b_{kl;s}}  
\end{eqnarray}
where $\beta=2$. Using both the equalities (15) and (16) in eq.(14), 
 we obtain the desired F-P equation  

\begin{eqnarray}
{\partial P\over\partial Y} &=& \gamma
\sum_{n}{\partial \over \partial\mu_n}\left(\mu_n P\right) 
+\sum_n {\partial \over \partial\mu_n}
\left[ {\partial \over \partial\mu_n} +
 \sum_{m\not=n}{\beta \over {\mu_m-\mu_n}}\right] P 
\end{eqnarray}

Here the left hand side of above equation, summing  over all  
$y_{kl;s}$ and $b_{kl;s}$, has been rewritten as 
${\partial P\over\partial Y}$ with $Y$ given by the condition that

\begin{eqnarray}
{\partial P\over\partial Y} 
=2\sum_s \sum_{k \le l} y_{kl;s}(\gamma - y_{kl;s})
{\partial P\over\partial y_{kl;s}} -   \gamma
\sum_s \sum_{k \le l} b_{kl;s}{\partial P\over\partial b_{kl;s}}.   
\end{eqnarray}
 By using the orthogonality of eigenvectors and 
following the same steps, it can be proved for 
real-symmetric  case too (now $\beta=1$)\cite{ps1}. It is worth noting  
that the eq.(17) is same as the 
evolution equation for the eigenvalues of Brownian ensembles.   
It is also similar to the one governing the transitions between any two 
universality classes of SGE caused by a random perturbation of strength 
$\tau$ (with $\tau \rightarrow Y$) \cite{bs,fh}. 

\subsection{How to Obtain the Complexity parameter $Y$:}

	The variable $Y$, a function of      
relative values of the coefficients $\alpha_{kl;s}$'s and $b_{kl;s}$'s, 
is a measure of the degree and nature of the complexity of a system 
 and  can therefore be referred as the "complexity parameter". 
For the case discussed here (eq.(18)), 
$Y$ can be obtained by the following method. 

We define 
$M=2 N^2$ variables $(Y_1,..Y_M)$ as the functions of all $y_{kl;s}$'s 
and $b_{kl;s}$'s  such that the condition given by eq.(18) 
(where $Y\equiv Y_1$) 
is satisfied. This is indeed possible by using the orthogonal (Jacobi) 
coordinate transformation between  variables $\{Y_i\}_{i=1,..,M}$ 
and $\{y_{kl;s}, b_{kl;s}\}_{k\le l; k,l=1,..,N; s=1,2}$  
defined by following rule,

\begin{eqnarray}
Y_i=  \sum_{j=1}^{M} a_{ij} X_j \qquad {\rm for}\quad i=1\rightarrow M  
\end{eqnarray}
where $X_j\equiv{1\over 2}{\rm ln} {y_{kl;s} \over |y_{kl;s}-\gamma|}+c_j$ 
for $j \le N^2$ and 
$X_j \equiv - {1\over \gamma}{\rm ln}|{b_{kl;s}}|+c_j$ 
for $j >N^2$ with $c_j$ as arbitrary 
constants of integration. Here coefficients $a_{ij}$ must 
satisfy the relation   
$\sum_{j=1}^M a_{ij}= \delta_{i1}$     
 which is  a necessary condition for orthogonality but not 
sufficient to get the right form for $\partial\over \partial Y$. With 
$D$ being the functional derivative of  ${Y_i}$'s  
with respect to $X_j$'s, we also need the elements $D^{-1}_{1j}$ of its inverse 
to be unity. One way to achieve this is to set all adjuncts of the matrix 
elements ${\partial Y_1\over \partial X_j}$ equal. Now by choosing $a_{1j}$
also equal, $a_{1j}= M^{-1}$, we are left with $M$ conditions 
for $a_{ij}$, $i\not=1$, which can easily be fulfilled.

The form of $Y = \sum_j a_{1j} X_j$, fulfilling condition (18), can therefore  
be given as 
\begin{eqnarray}
Y= {1\over 2 N^2} \sum_{k\le l} \sum_{s=1}^2 \left[ {1\over 2}
 {\rm ln}{y_{kl;s}\over |y_{kl;s}-\gamma|} -
{1\over \gamma} {\rm ln} |b_{kl;s}|\right] +C 
\end{eqnarray}
 with  
 $C= M^{-1} \sum_j c_j$.  

As obvious, this method is applicable only for the case when the prefactor 
associated with a derivative of $P$ with respect to a variable  $r$ in 
eq.(18) depends only on $r$
($r$ can be any one of the $y_{kl;s}$ or $b_{kl;s}$). 
Our studies on the ensembles 
 more complicated than eq.(8) show that the prefactors can also depend 
on a combination of various $r$ variables. This requires a more general 
method to 
obtain $Y$ which can also be used for the case discussed here (Appendix F).

\subsection {Determination of $P(\mu,Y)$}

The eq.(17) describes an evolution of the eigenvalues of GGE 
due to changing distribution parameters of the ensemble
 which can be solved, in principle, to obtain 
$P(\mu, Y)$ for arbitrarily chosen initial values of the  
parameters.  If the ensemble corresponding to initial set of the 
parameters is referred as $H_0$, an integration over  $H_0$ would 
lead to $P(\mu,Y)$, free of initial conditions. 
In fact, it can be shown that   

\begin{eqnarray} 
P(\mu;Y) &=& {(4\pi Y)^{-N^2 /2}} 
\int {\rm exp}\left[{-{1\over 4 Y}\;tr(\mu-U^+ \mu_0 U)^2}\right]f(\mu_0)
|\Delta(\mu_0)|^{\beta_0}{\rm d}\mu_0 {\rm d}U
\end{eqnarray}
   
where $\mu_0$ is the set of eigenvalues of the initial matrix $H_0$, with 
$\beta_0$ given by its symmetry conditions, and $U$ is the integral 
over unitary (or orthogonal) space of matrices.

To show that eq.(21) is indeed a solution of eq.(17), 
we study a general case. Consider a 
 partial differential equation for a function $F(A;t)$ defined in the 
matrix space of $N\times N$ Hermitian matrices $A$

\begin{eqnarray}
{\partial F \over \partial t} &=& \left[ 
\nabla^2_A F + \nabla.(A F)\right] \\
{\rm where}\qquad \nabla^2_A &=& \sum_i{\partial^2 \over \partial
 A_{ii}^2} 
+{1\over 2} \sum_{i<j} {\partial^2 \over \partial A_{ij}^2}
\qquad{\rm and}\qquad \nabla.(AF) = \sum_{i\le j} 
{\partial \over \partial A_{ij}} (A_{ij} F)
\end{eqnarray}

with the initial condition $F(A;0)=f(A)$. This equation is known to have 
a unique solution (see page 174 of \cite{meta}),

\begin{eqnarray} 
F(A;t) &=& \int K(A,B,t)f(B){\rm d}B \\
{\rm where} \qquad K(A,B,t) &=& {(4\pi t)^{-N^2 /2}} 
{\rm exp}\left[-{1\over 4t}\;
tr(A-B)^2\right] 
\end{eqnarray}

where $B$ is a $N\times N$ hermitian matrix.
Depending on the
nature of both $A$ and
$B$, we can choose a special class of eigenvector matrices 
$U_A$ and $U_B$ (for $A$ and $B$ real-symmetric, complex
hermitian or symplectic, $U_A$ and $U_B$ are orthogonal,
 unitary and symplectic matrices respectively) such that 

\begin{eqnarray} 
A=U_A^{s}a U_A \qquad {\rm and} \qquad B=U_B^{s}b U_B
\end{eqnarray}

where $a=[a_i\delta_{ij}]$,  $b=[b_i \delta_{ij}]$ are
diagonal matrices with $a_i$ and $b_i$ as the eigenvalues
of $A$ and $B$ respectively and $U^s=U^+$ or $U^T$ or $U^R$ depending 
on whether $U$ is an eigenvector matrix for a complex Hermitian, 
real symmetric or symplectic matrix \cite{meta}.
Let $\beta_A$ and $\beta_B$ give the number of components of a typical 
matrix elements in $A$ and $B$ respectively. 
 Changing the variables from
matrix elements to the $N$ eigenvalues and $\beta N(N-1)/2$ angle
(i.e eigenvector) parameters on which $U_B$
depends, we have 

\begin{eqnarray}
{\rm d}B = |\Delta(b)|^{\beta_b} \;{\rm d}b \;{\rm d}U_B 
\qquad {\rm with} \qquad 
{\rm d}b={\prod_{i=1}^N}{\rm d}b_i
\qquad {\rm and} \qquad
\Delta(b)=\prod_{i\le j} (b_i -b_j)  
\end{eqnarray}

The substitution of these relations  in eq.(24) gives us

\begin{eqnarray} 
F(A;t) &=& {(2\pi t)^{-N^2 /2}} 
\int {\rm exp}\left[-{1\over 2t}\;tr(a-U^s
bU)^2\right]f(b,U_B)|\Delta(b)|^{\beta_b}{\rm d}b {\rm d}U_B  
\end{eqnarray}

where $U=U_B U_A^s$ and $U^s=U_A U_B^s$. Now if $f(b,U_B)$ is independent
of $U_B$ then $F(A;t)$ is also independent of
$U_A$. 
This helps us to rewrite the eq.(28) as follows, 

\begin{eqnarray} 
F(a;t)  
&=&{(4\pi t)^{-N^2 /2}} 
\int G(a,b,t)f(b)|\Delta(b)|^{\beta_b}{\rm d}b
\end{eqnarray}

where 
\begin{eqnarray}
G(a,b,t)=\int {\rm exp}\left[{-{1\over 4t}\;tr(a-U^s bU)^2}\right]
{\rm d}U\nonumber
\end{eqnarray}

Here the integral is over the group $U$ of orthogonal, unitary and 
symplectic matrices respectively. 
Further the Laplacian $\nabla^2_A$ can also be written in
terms of eigenvalues and angle parameters of $A$ (see appendix A.5 of
\cite{meta}) 

\begin{eqnarray}
\nabla^2 (A) = {1\over|\Delta (a)|^{\beta_a}} \sum_i {\partial \over
\partial a_i}{|\Delta (a)|^{\beta_a}{\partial \over \partial a_i}}
+ \nabla^2_{U_A}.
\end{eqnarray}

By the substitution of eq.(30) in eq.(22) and using independence of
$F(a;t)$ of $U_A$, one can rewrite eq.(22) as follows,

\begin{eqnarray}
{\partial F (a;t)\over \partial t} = {1\over |\Delta (a)|^{\beta_a}} 
\sum_i {\partial \over
\partial a_i}\left[|{\Delta (a)|^{\beta_a}}
{\partial F (a;t)\over \partial a_i}\right] 
+ \sum_i {\partial \over
\partial a_i}(a_i F)
\end{eqnarray}

with $F(a;t)$ given by eq.(29). Now by using the equality 
$\sum_i {\partial^2 \over
\partial a_i^2}|\Delta (a)|^{\beta_a} = 0$, eq.(31) can be reduced in 
the following form: 

\begin{eqnarray}
{\partial F\over\partial t} &=& 
\sum_{i}{\partial \over \partial a_i }\left(a_i F \right) 
+\sum_i {\partial \over \partial a_i}
\left[ {\partial \over \partial a_i} +
\sum_{j\not=i}{\beta_a \over {a_j - a_i}}\right] F 
\end{eqnarray}

which is similar to eq.(17) with $a_i \rightarrow \mu_i$, $t \rightarrow Y$,
$\gamma=1$ and 
$ F \rightarrow P$. The   
 joint probability density $P$ can therefore be obtained by 
evaluating the integral (29). However, so far, the integration could be 
performed only 
for the unitary group of matrices \cite{itz,meta}.

\subsection{Steady State, Level Density and Correlations}

	The steady state of eq.(17), 
$P(\mu,\infty)\equiv P_{\infty}= |\Delta (\mu)|^{\beta}
{\rm e}^{-{\gamma\over 2}\sum_k \mu_k^2}$, 
corresponds to    $Y-Y_0 \rightarrow
\infty$ (with $Y_0$ as the complexity parameter of initial ensemble) 
which can be achieved by two ways (for finite $Y_0$ values). 
The first is when almost all 
$y_{kl;1} \rightarrow \gamma$   and $y_{kl;2} \rightarrow 
\infty$ (for finite $b_{kl;1}$ and $b_{kl;2}$ values ) 
which results in a GOE steady state. The second is when almost all  
$y_{kl;1} \rightarrow \gamma,\quad  y_{kl;2} \rightarrow \gamma$, resulting 
in a GUE. 
 This indicates that, in the steady 
state limit, system tends to belong to one of the SGEs. 
The eq.(17) can, therefore, describe a transition from a given initial ensemble 
(with $Y=Y_0$) to either GOE or GUE with $Y-Y_0$ as the transition parameter. 
The non-equilibrium 
states of this transition, given by non-zero finite values of $Y-Y_0$, 
are various Gaussian ensembles corresponding to varying values of 
the coefficients $y_{kl;s}$ and $b_{kl;s}$. 
For example, the choice of the initial 
ensemble as GOE (almost all $y_{kl;1} = \gamma, y_{kl;2} 
\rightarrow \infty$ initially) and a decrease of $y_{kl;2}$ 
(from $\infty \rightarrow \gamma$ while keeping $y_{kl;1}$ fixed) leads to 
GOE $\rightarrow$ GUE transition with 
intermediate ensembles as those of complex Hermitian matrices.      
Similarly Poisson $\rightarrow$ GUE transition can be brought about by 
choice of the initial ensemble as Poisson  
(almost all $y_{kl;1}, y_{kl;2} \rightarrow \infty$ 
for $k\not= l$, 
$y_{kk;1}=\gamma$, $y_{kk;2}=\gamma$ and $b_{kl;s} 
=0$ for all $k,l,s$ values) 
and by varying both $y_{kl;1}$ and $y_{kl;2}$ upto $\gamma$. 
As clear from above, $\gamma$ fixes the variance of the final ensemble 
and an arbitrariness in $\gamma$ leaves the latter arbitrary. This however 
does not affect the statistical 
properties of the intermediate ensembles.

The eq.(21) for $P(\mu,Y)$ can be used to obtain 
 $n^{\rm th}$ order density correlator $R_n (\mu_1,..\mu_n; Y)$, defined by 
 $R_n = { N! \over {(N-n)!}}\int P(\mu, Y) 
{\rm d}\mu_{n+1}..{\rm d}\mu_N$. 
($R_n$ can also be expressed in the form $<\nu (\mu_1,Y)..\nu(\mu_n,Y)>$
with $\nu(\mu,Y) = N^{-1} \sum_i \delta (\mu-\mu_i)$ as the density of 
eigenvalues  and $<..>$ implying the ensemble average).  
Here note that the analogy of eq.(17) 
to that of Dyson's Brownian ensembles implies same form of $P$ for both the 
cases and therefore $R_n$. A lot of information already being available about 
level-density and various correlation for Brownian ensembles, it can directly 
be used for ensemble described by eq.(8). Thus, as for BE, 
 a direct integration of F-P equation (17)  
leads to the BBGKY hierarchic relations among the unfolded local correlators 
$R_n(r_1,..,r_n;\Lambda)={\rm Lim} N\rightarrow \infty \;
{{\it R}_n(\mu_1,..,\mu_n;Y) \over 
{\it R}_1(\mu_1;Y)...{\it R}_1(\mu_n;Y)}$
with $r=\int^{r} {\it R}_1(\mu;Y){\rm d}\mu $ and 
$\Lambda=(Y-Y_0)/D^2$ ($D= R_1^{-1}$; the mean level spacing) \cite{ap},
 
\begin{eqnarray}
{\partial R_n \over\partial \Lambda} &=& 
\sum_j {\partial^2 R_n\over \partial r_j^2}
-\beta \sum_{j\not=k} {\partial \over \partial r_j} 
\left({R_n \over {r_j-r_k}}\right) 
-\beta \sum_j {\partial \over \partial r_j} 
\int_{-\infty}^{\infty} {R_{n+1} \over {r_j-r_k}}
\end{eqnarray}

(here for simplification, $\gamma$ is chosen to be unity).
As can be seen from the above equation, the transition for $R_n$ occurs 
on the scales determined by $Y \approx D^2$ and a smooth transition can 
be brought only in terms of the parameter $\Lambda$, obtained by rescaling 
$Y$ by $D^2$. On the other hand,  
for $R_1$, the corresponding 
scale is given by $Y \approx N\; D^2$. This implies, therefore, during the 
transition in $R_n$, the density $R_1$ remains nearly unchanged; this fact 
is very helpful in unfolding the correlators $R_n$.
For $n=1$ and in large $N$-limit, above equation reduces to the  
Dyson-Pastur equation \cite{ap} for the level density 
$<\nu(\mu_1,Y)> \equiv N^{-1} R_1$ 

\begin{eqnarray}
{\partial <\nu(\mu)>\over\partial Y} = 
-\beta {\partial \over \partial \mu}\left( 
\sum_m {\bf P}\int {\rm d}\mu'
{<\nu (\mu')> \over {\mu-\mu'}}\right) <\nu(\mu)> 
\end{eqnarray}

 which results in  
a semi-circular form for $\nu$; $\nu(r)={2\over \pi \eta^2}
(\eta^2 - r^2)^{1/2}$ with $\eta^2 = 4 N (1+ Y^2)$ \cite{fkpt}. 
The application of  
super-symmetry (SUSY) technique \cite{klh} to ensemble (8) 
 gave a similar result (also see section 4.3 of \cite{meta}).  
\section{Connection to Calogero Hamiltonian}

A similarity transformation followed by a Wick rotation converts the 
F-P equation into a self-adjoint form \cite{bs}. 
This 
can be seen as follows. The F-P equation, in general, can be expressed 
in a form  
\begin{eqnarray}
  {\partial |P_Y> \over\partial Y} = -P |P_Y> 
\end{eqnarray}
where  $ P$ is a F-P operator with non-negative eigenvalues. Here 
$|P_Y>$ is a general state of operator $P$ at "time" $Y$ and its projection in 
eigenvalue space can be obtained by the usual operation $P(\mu,Y)\equiv 
<\mu|P_Y>$ 
(with $\mu$ as  set of the eigenvalues).  Let 
$P(\mu,Y_0) \equiv <\mu|P_0>$ be the 
equilibrium probability.
One can further define 
a vector $<0|\equiv \int {\rm d}\mu <\mu|..$ satisfying $<0|P =0$ thus 
implying the conservation of probability in "time" $Y$ in this state  
(the ground state). 
 The F-P 
operator can now be hermiticized through a similarity transformation 
$S^{-1} P S = H$ where $S$ is Hermitian and invertible operator depending 
only on the eigenvalues. Thus the ground state condition must be given by 
$HS|0>=0$ (as $P^{+}|0>=0$). 
Let the effect of similarity transformation on the state $|P_Y>$ and $|P_0>$ is 
expressed by  $|\psi> = S^{-1} |P_Y>$ and $|\psi_0>=S^{-1} |P_0>$ respectively. 
 The similarity transformation of eq.(35) will then give the desired form  
 $ {\partial |\psi> \over\partial Y} = -H |\psi>$; the ground state 
$|\psi_0>$
must now satisfy the condition $H |\psi_0>=0$. The comparison of the two 
different 
forms of the ground state condition gives  $|\psi_0>=S|0>$ and therefore 
$|P_0>= S^2 |0>$.

	In the case of F-P equation (17),
$H$ turns out to be CM Hamiltonian (eq.(1) with $r_i \rightarrow \mu_i$) 
 and has well-defined eigenstates and eigenvalues \cite{cal,znc}. 
As well-known, 
the particles in CM system undergo an integrable dynamics, thus implying 
a similar motion for the eigenvalues too. Here $H$ being a generic member of 
GGE, this result is valid for all systems with interactions complicated 
enough to be modeled by GGE.

The "state" 
$\psi$ or $P(\mu,Y| H_0)$ can be expressed as a 
sum over the  
eigenvalues and eigenfunctions  which on integration over  the 
initial ensemble  $H_0$ 
leads to the joint probability distribution $P(\mu,Y)$ and 
thereby static (at a single parameter value) density correlations $R_n$. The  
above correspondence can also be used to map the multi-parametric 
correlations  to multi-time correlations of the of CM Hamiltonian. For 
example,  
	 the parametric 
correlation $<Q_a(Y)Q_b(0)>$,  
for a classical variable $Q(Y)$ with $[Q,S]=0$ can be mapped to the 
corresponding ground state correlation of CM hamiltonain 
 $<\psi_0| Q_a(Y) {\rm e}^{-YH} Q_b(0) |\psi_0>$. 
This follows because 
\begin{eqnarray}
<Q_a(Y)Q_b(0)>   
=\int Q_a Q_b P(\mu,Y) {\rm d}\mu  
= \int <\mu|Q_a Q_b|P_Y>  {\rm d}\mu  
\end{eqnarray} 
now as the evolution of $|P_Y>$ with respect to $Y$ is given by 
$|P_Y> = S{\rm e}^{-Y H} S^{-1} |P_0>$, one has 
\begin{eqnarray}
<Q_a(Y)Q_b(0)>  = 
  <0|
Q_a S {\rm e}^{-Y H} S^{-1} Q_b |P_0> 
=  <\psi_0|Q_a {\rm e}^{-Y H} Q_b |\psi_0> 
\end{eqnarray}

\section{Application to Physical Problems}: 

        The given ensemble (8), referred here as "G", is represented 
by a point $Y$ in the parametric-space 
consisting of distribution parameters and various transition curves may pass 
through this point. 
The question therefore arises which curve should be chosen for 
the studies of the properties of $G$? The answer is the one which does not 
leave any arbitrariness behind and if there are more than one such curve, 
each one 
of them should give same answer for various fluctuation measures of $G$. 
This criteria for the right choice are  
based on the symmetry properties of ensemble $G$, that is, the 
nature of all $\alpha_{kl}$ and $b_{kl}$ with end-points 
(the final and initial ensemble, referred here as "F" and "O" respectively) 
chosen in such a way that the values corresponding to G occur during 
the variation of distribution parameters from one end to the other. 
Further the chosen transition should preferably be the one whose properties 
are already known and can therefore tell us about G. For many GGE 
described by eq.(8), above criteria is satisfied by choosing F as 
a SGE with variance $<F_{ii}^2>= 2<F_{ij}^2>=\gamma^{-1}$, 
$\gamma \le {\rm min}\{ y_{kl;s}[G]\}$, 
$k,l= 1,2,..,N$, $s=1,2$,  
and O as an ensemble with each $\alpha_{kl}$[O] given by the maximum 
value taken 
by the functional form of the corresponding $\alpha_{kl}$[G]. However, 
as explained in following examples, O can 
also be chosen as some other ensemble. 
For example, If G is an ensemble of real-symmetric matrices $H$ 
represented by 
$\rho(H) \propto {\rm exp}[-\sum_{k \le l} \alpha_{kl} H_{kl}^2]$ with finite 
but different values 
for all $\alpha_{kl}$, the Poisson $\rightarrow$ GOE curve is more suitable 
for its study rather than GOE  $\rightarrow GUE$.   
 Here the GOE ensemble is described by 
$<F_{ii}^2> =2 <F_{ij}^2> = \gamma^{-1}$ with $\gamma$ as the 
minimum value among  
given $y_{kl}[G]$s. 
However if  various $\alpha_{kl}$ in the above example can also 
take complex values, 
the ensemble can now be chosen on any one of the curves, namely, Poisson 
$\rightarrow$ GUE or GOE $\rightarrow$ GUE. Here now GUE can be chosen as    
 $<F_{ii}^2> = 2 <F_{ij;1}^2> = 2 <F_{ij;2}^2>=\gamma^{-1}$. 
  The GOE for the second curve 
 can be chosen as the  one with 
 $<O_{ii}^2> = 2 <O_{ij;1}^2> =q^{-1}$ and $<O_{ij;2}^2>=0$
 with $q={\rm max} \{y_{ij;1}[G]\}$.
Similarly, for Poisson $\rightarrow$ GUE curve, the initial ensemble 
may be taken as one with 
 $<O_{ii}^2> = \gamma^{-1}$ (or $q^{-1}$) 
and $<O_{ij;1}^2> = <O_{ij;2}^2>=0$ for 
$i\not= j$. 
The reason for the choice of the two transitions is due to  
availability of 
the results for their 2-point correlation $R_2$ \cite{ap}: 

For Poisson $\rightarrow$ GUE

\begin{eqnarray}
 R_2 (r;\Lambda) - R_2(r;\infty)={4\over \pi}\int_0^\infty {\rm d}x 
\int_{-1}^1 {\rm d}z \;{\rm cos}(2\pi rx) 
\;{\rm exp}\left[-8\pi^2\Lambda x(1+x+2z\sqrt x)\right]
\left({\sqrt{(1-z^2)}(1+2z \sqrt x) \over 1+x+2z \sqrt x}\right)
\end{eqnarray}

and for GOE $\rightarrow$ GUE

\begin{eqnarray}
 R_2 (r;\Lambda) - R_2(r;\infty)=-{1\over \pi^2}\int_0^\pi {\rm d}x 
\int_{\pi}^{\infty} {\rm d}z \; x\;{\rm sin}(rx) 
\;{\rm exp}\left[2\Lambda (x^2 - y^2) \right]\; 
{{\rm sin}(y r)\over y}
\end{eqnarray}

where $R_2(r,\infty)=1-{{\rm sin}^2(\pi r)\over \pi^2 r^2}$ (the GUE limit). 

It is obvious therefore that if $\Lambda_1$ and $\Lambda_2$ are the 
parameter values for the ensemble "G" on Poisson $\rightarrow$ GUE 
and GOE $\rightarrow$ GUE curves respectively, one should have 
$R_{2, P\rightarrow U} (r;\Lambda=\Lambda_1) =
R_{2, O\rightarrow U} (r; \Lambda=\Lambda_2)$.
This would require an intersection of two curves in the $R_2 - \Lambda$  
space which however is possible. This is because the GOE can occur 
as an intermediate 
point in Poisson $\rightarrow$ GUE transition. The GOE $\rightarrow$ GUE 
curve can also appear as a part of the Poisson $\rightarrow$ GUE curve;   
thus the  
choice of two different initial ensembles here corresponds only to 
two different origins of dynamics on the same curve.  

The parameter $\gamma$, which determines $Y$ as well as the variances of $F$,  
enters in calculation at step given by eq.(14) and 
can be chosen arbitrarily. 
As suggested by  eq.(17), the choice of different $\gamma$-values  
corresponds to different $Y$-values as well as the transition
curves with end-points of same nature but
different variances; 
  this, however, would not imply different 
properties for the ensemble G (Appendix G). 
Similarly the F-P equation is although independent of the choice of the 
initial ensemble, the latter is required for determination of 
the correlations of G. 
The possibility of an arbitrary choice of O may seem to imply 
a certain arbitrariness left in the correlation of G. However the choice of 
two different initial ensembles corresponds only to the two different origins 
of the dynamics approaching to the same point in the parametric space.

It will be clarified by the examples given below.

\subsection{Anderson Transition}:

Using above method, 
the transition parameter for a metal-insulator transition as a result of 
increasing disorder can exactly be calculated. To see this, let us consider 
the case of a d-dimensional disordered lattice, of size $L$, 
in tight-binding approximation. Here, in the 
configration space representation of the Hamiltonian, a $N\times N$ matrix 
of size $N=L^d$, the diagonal 
matrix elements will be site-energies $\epsilon_i$.  The hopping is generally 
assumed to connect only the $z$ nearest-neighbors with amplitude $t$ so that 
the electron kinetic energy spread or bandwidth is $zt$. This therefore 
results in sparse form of the matrix $H$. We first consider 
the case of $L\rightarrow D$ transition brought about by decreasing diagonal 
disorder only. In this case, site-energies $\epsilon_i$ 
are taken to be independent random variables with 
probability-density $p(\epsilon_i)$. In the Anderson model \cite{and} 
of metal-insulator 
(MI) transition, $p(\epsilon)$ was taken to be a constant $W^{-1}$ between 
$-W/2$ to $W/2$. Various physical arguments and approximations used in this 
case led to 
conclusion that here all the states are localized for 
$W > 4 Kt {\rm ln}({W\over 2 t})$ with $K$ as a function of $z$ and $d$.   

However, as  well-known now,  MI transition does not depend 
on nature of $p(\epsilon)$ and latter can also be chosen as Gaussian; 
 the type of $p(\epsilon)$ affects only the critical point of the transition.  
The $\rho(H)$, for any intermediate state of MI transition brought about  
by diagonal disorder, can therefore be chosen 
as in eq.(8) with 
$\alpha_{kl}\rightarrow \infty$, $b_{kl}=-t$ for $k\not= l$, 
$\alpha_{kk}=\alpha$ and $b_{kk}=0$ for all $k$ which results in        
$Y  = {1\over 2 N^2}\left[ {N\over 2}{\rm ln} 
{2\alpha\over |2\alpha-\gamma|} - \gamma^{-1} K {\rm ln} t \right] + C$. 
 Here $K$ is total number of the sites connected and 
depends on the 
dimensionality $d$ of the system.
The system can initially be considered in an insulator regime where all the 
eigenvectors become localized on individual sites of the lattice 
(strong disorder limit). This results in a diagonal form of the matrix $H$ 
  with the eigenvalues  
independent from each other. The insulator limit can therefore be modeled 
by ensemble (8) with $\alpha_{kl} \rightarrow \infty$ for $k\not= l$, 
 $\alpha_{kk}=\alpha_0$ (for all $k$-values) 
 and $b_{kl}\rightarrow 0$ (for all $k,l$), giving, 
$Y_0 = {1\over 4 N}{\rm ln} 
{2\alpha_0\over |2\alpha_0-\gamma|} + C$ (as K=0 in the insulator regime). 
 The decrease of the diagonal disorder, that is,  
 an increase of $\alpha_{kk}$ from $\alpha_0$ to some finite values
(while $\alpha_{kl}$, $k\not= l$, remains infinite throughout the transition)
 will ultimately lead to metal regime with 
fully delocalized wavefunctions. The eigenvalue distribution of $H$ in the  
regime can be well-modeled by the SGE; let it be 
described by  $\alpha_M$($ > \alpha_0$). Thus for the study of transition 
in this case 
we should choose $\gamma=2\alpha_M$.   
The transition parameter can now be given as follows, 
with the mean level spacing $D\propto {1 \over \sqrt N}$,  

\begin{eqnarray}
\Lambda = {Y - Y_0\over D^2} = {1\over 4 }\left[{\rm ln} 
{\alpha |\alpha_0 -\alpha_M|\over \alpha_0|\alpha-\alpha_M|} - 
{K\over N \alpha_M} {\rm ln}\; t
 \right]
\end{eqnarray}
As obvious from the above, the transition is governed  
by relative values of the disorder and the hopping. Here 
$\Lambda \rightarrow 0$  leads to fully localized regime which 
corresponds to following condition on $\alpha$ and $t$ 
\begin{eqnarray}
{\rm ln} {\alpha \over \alpha_0} + {\alpha -\alpha_0 \over \alpha_M}  
&=&  {K\over N \alpha_M}{\rm ln}\; t
\end{eqnarray}
The eq.(40) gives, therefore, the condition for the 
critical region or mobility edge (${K\over N}\rightarrow$ finite 
as $N \rightarrow \infty$). 
As ${|\alpha -\alpha_0|\over \alpha_M} << 1$ even for large $\alpha$-values,  
 the condition is always satisfied if  
${K\over N\alpha} \rightarrow 0$. This explains the  localization of all the  
states in infinitely long wires (or strictly 
1-d systems where $K << N$) even for very weak 
disorder.  With increasing dimensionality $d$, connectivity $K$ of the lattice 
 and thereby the possibility of $|\Lambda| >> 0$ and the delocalized
 states increases. The $\Lambda$ can 
similarly be calculated when off-diagonal disorder is also present.

\subsection{1-D, Quasi 1-D, Periodic 1-D disordered and Chaotic systems}:

	In 1-D geometry of a solid state system e.g a chain of $N$ interacting 
sites, in tight binding approximation, the long-range random hopping leads to 
a banded structure of the matrix, known as random banded matrix (RBM) 
\cite{izra,fm2}. 
Here the effectively  non-zero, randomly distributed, matrix 
elements are  confined within a band with its width governed by the range of 
hopping. 
The 1-D periodic geometry e.g. a chain of interacting sites joined into a ring 
leads to a periodic RBM in which all non-zero matrix elements belong to 
three regions: a band along the main diagonal, the upper right corner and 
the lower left one \cite{izra}.
	A real disordered wire has finite cross-section (referred as 
quasi 1-D geometry) and therefore allows 
for propagating modes with different transverse quantization numbers frequently 
 referred as transverse channels. This case can again be modeled by RBMs with 
band-width given by number of transverse channels \cite{fm91,fm2}.
      In the case of dynamical systems too, 
exhibiting strong chaos in classical 
limit, the generic structure of Hamiltonian matrix in some basis is banded and 
matrix elements can be assumed to be pseudo-random \cite{pr}. 
For example, the Hamiltonian of quantum kicked rotor turns out to be 
a RBM in momentum basis \cite{izra}. 

In all these cases, nature of the disorder or associated randomness 
decides the  
nature of the distribution of 
the matrix elements. The physical properties of such systems can therefore be 
analyzed by studying the distribution of the eigenvalues of associated RBMs. 
      The most studied type of RBM is that with the zero mean value of all 
matrix elements and variance given by $<H_{nm}^2> = v^2 a(|n-m|/b)$ 
where $a(r)$ is some function satisfying the condition 
${\rm lim}_{r \rightarrow \infty} a(r) =0$ and determines the shape 
of the band \cite{fm2}. 
For large but finite size of the matrix $N >> b >>1$, its statistical 
properties were shown (by SUSY method) to be determined by the 
scaling parameter $b^2/N$  
with  the transition parameter  
 scaling as $Nf({b^2\over N})$ \cite{izra}.   

	The transition parameter for the RBM can also be calculated by 
our method.  
Let us first consider the simplest case i.e. Rosenzweig-Porter Model where
all the off-diagonal matrix elements are distributed with same variance  
which is different 
from the diagonal ones.  Let us take 
$\alpha_{ij; i\not=j}[G]= 2(1+\mu)$ and $\alpha_{ii}[G] = 1$ with 
$\mu \ge 0$; thus min$\{y_{ij}[G]\}=2$ and we can choose $\gamma =2$.  
This GGE can therefore be mapped to a Brownian ensemble, with  
$Y-Y_0 =- {N-1 \over 4 N} {\rm ln} |1- {1\over 1+\mu} | 
\approx {1 \over 4 \mu}$ for $\mu > 1$, 
appearing in a Poisson $\rightarrow$ GOE 
transition where the initial matrix elements distribution is given by 
 $P(H_0)\propto {\rm e}^{-\sum_i H_{ii}^2}$ and the final, stationary state, 
 obtained for large $\Lambda$-values,  is 
 $P(H)\propto {\rm e}^{-{\gamma \over 2} {\rm Tr} H^2}$.   
Now as $R_1 \approx \sqrt{N}$ \cite{fh,klh}, 
the $D^2 \approx 1/N$ and therefore 
$\Lambda \approx {N \over 4 \mu}$ which implies that the GGE will have an 
eigenvalue statistics very different from that of Poisson or GOE only if 
 $\mu \approx c N$ ($c$ a finite constant). 
For $\mu > c N$, $\Lambda \rightarrow 0$ 
and  for $\mu < c N$, $\Lambda \rightarrow 
\infty$ for $N\rightarrow \infty$ and  thus   
 the GGE behaves like a Poisson ensemble in the first case and like a 
GOE in the second; 
this result 
is in agreement with the one obtained, in \cite{ajs}, by using NLSM technique. 
(Note in ref. \cite{ajs}, $D$ is taken as $D \propto 1/N$, 
which gives $\Lambda \approx {N^2 \over 2\mu}$ and therefore GOE and Poisson 
ensemble result for $\mu < c N^2$ and $\mu > c N^2$ respectively). 
 
 
..

Consider the ensemble with exponential decay of the variances 
away from the diagonal 
i.e $\alpha_{kl} = {\rm e}^{|k-l|/b}, k\le l, 1 << b << N$. 
 Thus, again $\gamma=2$ and the final ensemble is a SGE with 
$P(H)={\rm e}^{-{\gamma\over 2} {\rm Tr} H^2}$ 
and therefore 
$Y=-{1\over 4 N^2} \sum_{i\le j=1}^N 
{\rm ln} |1- \gamma  y_{ij}^{-1}| + C$. 
 Here the initial ensemble is 
that of the diagonal matrices with  a Poisson distribution of 
the eigenvalues which  corresponds to 
  $y_{ii}$[O]$=2$  and 
$y_{ij; i\not=j}$[O] $\rightarrow \infty$ (this being maximum value 
of $y_{kl}$[G]) giving $Y_0  
=-{1\over 4 N^2 } \sum_{i=1}^N {\rm ln}
 |1- \gamma  y_{ii}^{-1}{\rm [O]}| + C $.  
 Thus 
$Y-Y_0=-{1\over 4 N^2} \sum_{r=1}^N (N-r){\rm ln} |1- 2{\rm e}^{-r/b}| 
\approx {b \over N}$.   
As $R_1 \approx \sqrt{N}$, the   
transition parameter for infinite system ($N\rightarrow \infty$) turn out  
to be $\Lambda = {Y\over D^2} \approx b$ (see \cite{ps1}) 
which reconfirms that, in infinite 
systems, the transition is governed only by the band-width  $b$
\cite{fm2,izra}. 

 Another case of importance is the ensemble with  power law decay of variances
$H_{ij}=\tilde G_{ij} a(|i-j|)$ with 
$\tilde G$ a typical member of SGE $(<\tilde G_{ii}^2>= 
2<\tilde G_{ij}^2>=v^2)$
and $a(r)=1$ and $(b/r)^\sigma$ for 
$r\le b$ and $r > b$ ($b>>1)$ respectively (known as PRBM model with 
P stands for power) \cite{mfdqs}. This corresponds to 
$y_{ij}= {1 \over v^2 a^2(|i-j|)}$ and therefore 
$\gamma ={\rm min} \{y_{ij}\}= {1\over v^2}$.  
Again as for the exponential case, the choice of initial and the final ensemble 
   remains the same.     
Now as   $y_{ij}\equiv y_r=\gamma ({r\over b })^{2\sigma}$ 
(with $r=|i-j|$), we get  
\begin{eqnarray}
\Lambda = D^{-2} (Y-Y_0) &=& -{1\over 4 N} 
 \sum_{r=b+1}^N (N-r) 
{\rm ln} \left( 1 - \left({b\over r}\right)^{2\sigma}\right)\\ 
&\approx & {N\over 4} 
\sum_{j=1}^{\infty}{1\over j} 
 \left({b\over N}\right)^{2 j \sigma} 
\int_{b/N}^1 {\rm d}x (1-x) x^{-2j\sigma} \nonumber \\
&=& {N\over 4}\sum_{j=1}^{\infty}{1\over j} \left[
{1\over 2 (1-2 j\sigma)(1-j\sigma)}\left({b\over N}\right)^{2j\sigma} 
- {1\over  (1-2j\sigma)}{b\over N} 
+ {1\over 2 (1-j\sigma)}\left({b\over N}\right)^2 \right] 
\end{eqnarray}
Thus, for large $N$-values and $\sigma < 1/2$, 
$\Lambda (\propto N^{1-2\sigma})$ is 
sufficiently large and the 
eigenvalue statistics approaches SG limit. 
At $\sigma=1/2$, the statistics is governed by 
the parameter $b^2/N$ instead of $N$ only.  
For $ \sigma =1$, 
 the 
non-zero, finite $\Lambda$-value ($\Lambda \propto b$ 
even when $N \rightarrow \infty$) 
leads to an eigenvalue 
statistics intermediate between  
 that of SGE or Poisson. 
 For $\sigma > 3/2$ with $N\rightarrow \infty$, $\Lambda \rightarrow 0$ 
 therefore, the 
eigenvalue statistics approaches Poisson limit, $\Lambda$ being very small. 
 All these results are in 
agreement with those obtained 
in \cite{mfdqs} by SUSY technique. 

      Another type of RBMs often encountered in atomic and nuclear systems 
 are those with the non-zero mean value of all 
matrix elements and with variance given by $<H_{nm}^2> = v^2 a(|n-m|/b)$; 
the transition parameter for them can also be obtained 
as for the above cases \cite{fggk,wig,fcic,pr,shep}. 

\subsection{Quantum Hall Case}:

 A Quantum Hall system without disorder 
has all the  states degenerate within a single Landau level. 
The introduction of 
disorder leads 
to a broadening of the levels (also termed as diagonal disorder) as well 
as random hopping between them (off-diagonal disorder) 
and a competition between 
the two leads to a L $\rightarrow D$ transition. Note this is different from 
 the 
Anderson model where the L$\rightarrow D$ transition is caused by the 
competition 
between diagonal disorder and non-random hopping (bandwidth) \cite{and}. 
The $N\times N$ Hamiltonian matrix in presence of disorder therefore belongs 
to an ensemble far more complicated than eq.(8), known as 
random Landau matrix, as now various matrix 
elements are no longer independently distributed: 
 $\rho (H,y,b)=C{\rm exp}[ 
- \sum_{s=1}^2 \sum_{k,l; k\le l} H_{kl;s} (\alpha_{kl;s}
 H_{kl;s} - \sum_{i,j; i \le j}' b_{ijkl;s} H_{ij;s})] $   
 with $C$ as the normalization constant  
and $y$ and $b$ as the sets  
 of inverse of variances 
$y_{kl;s}={\alpha_{kl;s}g_{kl} }$ and coefficients 
$b_{ijkl;s} $ respectively  
with $g_{kl}=1+\delta_{kl}$. Here $\sum_{i,j}' b_{ijkl;s}$ will imply that 
the summation is over all possible pairs of indices $\{i,j\}$ such that 
the pair $\{i,j\} \not= \{k,l\}$ or $\{l,k\}$ \cite{bh}. 
In this case too, one can show that the eigenvalue distribution $P$ 
satisfies eq.(17) but the condition for the determination of $Y$ is 
no longer given by eq.(18); the details will be presented elsewhere. 

\subsection{Critical Ensemble and Multifractality of Eigenvectors}

	Recent studies of some metal-insulator transitions revealed that the 
energy level statistics in the critical region is universal and 
different from both Wigner-Dyson as well as Poisson statistics. 
The eigenfunctions associated with the critical statistics show 
multifractal characteristics \cite{bh1,klaa,km}.  
The level 
number varaince $\Sigma^2 (N)$ is believed to be an  
important indicator of this critical behaviour with its asymptotic 
linearity in the mean number of levels ${\bar N}$ \cite{ckl};  
$\Sigma^2 ({\bar N})= <(\delta N)^2> =\chi {\bar N}, \chi < 1$.
The critical statistics, therefore,  governs the 
spectral fluctuations that are weaker than for the Poisson statistics
($\Sigma^2({\bar N}) = {\bar N}$) but much stronger than for the Wigner-Dyson 
statistics,
($\Sigma^2({\bar N}) = {\rm ln}{\bar N}$). 
Later on remarkable similarities were 
found between the spectral statistics of a number of dynamical systems 
e.g pseudointegrable billiards and the critical statistics near the mobility 
edge \cite{bog}. However such a critical region being inaccessible either 
perturbatively or 
semiclassically, a different tool was required to probe into it. This 
led to the suggestion of a  
RM modelling of this region \cite{km}. The $N\times N$ matrices 
in this model are Hermitian and matrix elements are  Gaussian 
distributed with zero mean and the variance given by 

\begin{eqnarray} 
<(H_{ij})^2> = \left[ 1 + \left({|i-j| \over B}\right)^{2 \sigma} 
\right]^{-1}  
\end{eqnarray} 
Using SUSY technique, it has been shown \cite{mfdqs} that for large $B$-values 
($B>>1$), this ensemble behaves like a SGE for $\sigma <1$ and as a Poisson 
for $\sigma >1$. The case $\sigma=1$ is believed to be of special 
relevance as it 
supports critical statistics and multifractal eigenstates; the  
application of SUSY technique  gives   
$R_2(r) \approx  1 - {\frac{1}{16 B^2}}  
   {\frac{{\rm sin}^2(\pi r)}{{\rm sinh}^2(\pi r/4 B)}}$  
and $\Sigma^2 (N) \approx \chi N$ \cite{mcin,mns,km}.
 
 
	The existence of the ensembles with critical statistics is  
indicated by our technique too. 
	The $N$-dependence of the transition parameter $\Lambda$, 
entering through $Y$ and the mean level-spacing $D$,  causes 
the transition to reach the equilibrium in limit $N\rightarrow \infty$ for 
finite, non-zero $Y$-values. In some cases, however, the $N$-dependence 
of $Y$ may be such that it balances the one due to $D$, thus resulting in 
an $N$-independent $\Lambda$ (as shown in section III.A,B) 
and therefore critical statistics. As can be seen from eq.(20), $\Lambda$ 
for the ensemble, given by eq.(44), is also $N$-independent for 
$\sigma=1$; here the ensemble appears as an intermediate point between 
Poisson $\rightarrow$ GUE transition with  
$Y-Y_0 = {1\over 4 N^2} \sum_{r=1}^N (N-r) 
{\rm ln} \left( 1 + ({b\over r})^{2\sigma}\right)$ and  
$\Lambda$ behaves as in the case of PRBM model discussed above, 
showing criticality for $\sigma=1$.
The correlation $R_2$ for the ensemble (44) can therefore be given by   
 eq.(38) which for large $\Lambda$-values 
(for all $r$), 
can be approximated as follows \cite{ks,fgm}: 

\begin{eqnarray}
R_2(r,\Lambda) &=& 1 + {\frac {\Lambda}{\pi^2\Lambda^2+r^2}} + 
{\frac {1}{2\pi^2 r^2}} 
[{\rm cos}(2\pi r) {\rm e}^{- 2 \frac {r^2}{\Lambda}} - 1]\\
&=& 1 +\frac{1}{\pi^2 \Lambda} + \frac{1}{2\pi^2 r^2} 
[{\rm e}^{- 2 \frac{r^2}{\Lambda}}-2{\rm e}^{- 2 \frac{r^2}{\Lambda}} 
{\rm sin}^2 (\pi r) -1]\\ 
&\approx &  1 +\frac{1}{\pi^2 \Lambda} 
- \frac{{\rm sin}^2{\pi r}}{\pi^2 r^2 
{\rm e}^{2\frac{r^2}{\Lambda}}}
\approx 1 - { \frac{6}{\pi^2 \Lambda}}  
   \frac{{\rm sin}^2(\pi r)}{{\rm sinh}^2(r\sqrt{6}/\Lambda)}
\qquad ({\rm for}\quad r << \sqrt\Lambda)
\end{eqnarray} 
which is similar to the result given by SUSY technique. However,  
 for $\Lambda >> r >> {\sqrt\Lambda}$, our method gives 
$1 - R_2 (r,\Lambda) = 
 - {\frac {\Lambda}{\pi^2\Lambda^2+r^2}} + 
{\frac {1}{2\pi^2 r^2}}$ 
  while SUSY technique 
gives $1-R_2$ as an exponentially decaying function. 

As obvious from eq.(47), $R_2$ approaches GUE limit as $\Lambda 
\rightarrow \infty$ but, for finite $\Lambda$-values, it is very different 
from both Poisson as well as GUE. This indicates that the 
 ensembles with distribution parameters 
giving rise to a finite $\Lambda$ do not reach stationarity even for infinite 
size of their matrices, and, their properties being very different from those 
of the equilibrium ensembles, can be referred as "critical". However 
in our technique, as 
shown in previous sections,  the difference between various GG ensembles 
(within same stationarity limits) 
manifest itself only in different $\Lambda$-values, leaving the functional 
form of various statistical measures unaffected. Thus   
RP model as well as ensemble (44), both being GGEs and lying on 
Poisson $\rightarrow$ GUE curve,  would  
follow similar formulations for various statistical measures; 
For example,  $R_2$ for both of them is given by eq.(47) although with 
different formulas for $\Lambda$ and 
 both can show the critical behavior. However a contradiction 
arises when one considers the Number variance statistics $\Sigma^2 (r)$ 
which can be expressed in terms of $R_2(r)$ \cite{meta}, 
\begin{eqnarray}
\Sigma^2(r;\Lambda)= r - 2 \int_0^r (r-s)(1-R_2(s))\; {\rm d}s
\end{eqnarray}  
and therefore should show a similar behavior, as a function of 
$\Lambda$, for both (RP model 
and ensemble (44)).
But a detailed study of RP model by SUSY technique\cite{fgm} suggests that 
although it shows critical statistics for $\mu =c N$, 
it can not support 
linear nature of  $\Sigma^2=\chi r$ with $\chi <1$. As claimed by this study, 
the difference in $\Sigma^2 (r)$ behavior arises due to difference in 
large-$r$, $(\Lambda >> r >> {\sqrt\Lambda})$, behavior of $R_2(r)$ in 
the two cases.  

	As our technique is equally well-applicable to both these 
systems,  multifractality should exist in either both 
or none of them. Note that the multifractal nature of an ensemble is 
so far believed to be indicated by its $\Sigma^2$-behavior. But the 
latter is not yet clearly understood for RP model (see \cite{fgm,ks}) 
and therefore 
question of multifractality is still not fully settled. Also note that 
the earlier results for both models are obtained by SUSY technique 
using saddle point approximation at various stages which may also 
be misleading. It is also possible that (i) the seeming multifractality 
of ensemble (44) is  the erroneous conclusion of various approximations, 
 or (ii) $\Sigma^2 (r) \approx \chi r$ is not always 
a correct indicator of      
multifractality and therefore its absence in RP model. 

We believe that the $\Sigma^2(r)$-behavior is a bigger suspect 
 \cite{klaa,mir}. Our belief has its roots in the 
 direct applicability of our technique to 
Anderson model too.
Here also the ensemble for $H$ is 
located between Poisson $\rightarrow$ GUE (for a time-reversal breaking 
disorder) with corresponding $R_2$-behavior  given by 
eq.(38). Thus for finite $\Lambda$-values corresponding to critical region, 
the eigenvalue statistics is different from Poisson or GUE. But 
again for $\Sigma^2$ obtained by using  eq.(38), $\Sigma^2(r) \not = \chi r$ 
with $\chi <1$ and therefore 
if it is indeed an indicator of multifractality of eigenfunctions, 
our technique would suggest its absence in Anderson model. However the 
existence of multifractality among the eigenfunction of Anderson Hamiltonian 
is experimentally confirmed.

Our results indicate that multifractality will either be 
a common feature of all the Gaussian ensembles with finite $\Lambda$-values 
in limit $N\rightarrow \infty$ or it does not exist in any of them.  
Thus the questions related to critical 
statistics, the correct criteria for multifractality 
 and its analysis by SUSY technique require further probing.
\section{Other Cases}

\subsection{A perturbed Hamiltonian with GG type perturbation}

	The intimate connection of RMT $\rightarrow$ CM Hamiltonian 
continues also for system  $H=H_0 + x V$ 
with a random perturbation 
$V$ drawn from 
a GGE (i.e    
 $\rho (V,y,b)=C{\rm exp}({ 
- \sum_{s=1}^2 \sum_{k\le l} \alpha_{kl;s} (V_{kl;s}-b_{kl;s})^2})$. 
In this case, the 
eigenvalues evolve due to changing strength of perturbation too.   
To obtain the desired 
evolution equation, therefore, one needs to consider 
 the sum 
${\partial P\over\partial x}
 + \sum_s \sum_{k \le l}\left( \gamma - 
y_{kl;s}\right)\left[ 2 y_{kl;s} 
{\partial P\over\partial y_{kl;s}}
-b_{kl;s}{\partial P\over\partial b_{kl;s}}\right]$
 which leads to following equality 

\begin{eqnarray}
{\partial P\over\partial x } +   
\sum_s \sum_{k \le l}\left(\gamma - y_{kl;s}\right)\left[2 y_{kl;s}
{\partial P\over\partial y_{kl;s}}  
- b_{kl;s} {\partial P\over\partial b_{kl;s}} \right]  
=\sum_s \sum_{k\le l}  I_{kl;s} -
\sum_s \sum_{k\le l} y_{kl;s} I_{kl;s}
\end{eqnarray}

where $I_{kl;s}$ is still given by same form as eq.(12) 
but with $H$ replaced by $V$. As the right hand side of eq.(49) is same as 
that of eq.(14), one again obtains  
 obtains the evolution equation 
(17) but now $Y$ is given by the condition that 
${\partial P\over\partial Y } =    
{\partial P\over\partial x } +   
\sum_s \sum_{k \le l} \left(\gamma - y_{kl;s}\right) \left[ 2 y_{kl;s}
{\partial P\over\partial y_{kl;s}}  
- b_{kl;s} {\partial P\over\partial b_{kl;s}} \right]$. Proceeding just as 
in section I.C, $Y$ can be shown to be given by the following relation: 
\begin{eqnarray}
Y= {1\over 2 N^2+1)} \left[x + \sum_{k\le l} \sum_{s=1}^2 \left( {1\over 2}
 {\rm ln}{y_{kl}^{(s)}\over |y_{kl}-\gamma|} - \gamma^{-1}
{\rm ln} |b_{kl;s}|\right)\right] +C 
\end{eqnarray}

Again the steady state is achieved for $Y \rightarrow
\infty$ which corresponds to $x \rightarrow \infty$ and
$y_{kl;s} \rightarrow \gamma $; the steady state solution for $P$ is given by  
$\prod_{i<j} |\mu_i-\mu_j|^{\beta}
{\rm e}^{-{\gamma \over 2}\sum_k \mu_k^2}$. 
Here  only $x \rightarrow 
\infty$ 
(with ${\partial P\over \partial x}=0$ and $H=x V$) 
no longer represents a  
steady state, as in the case when V belongs to SGE,   
but represents a 
transition state 
with ${\partial P\over \partial Y} \not=0$. Note from eq.(50) that 
$Y\rightarrow \infty$ as $x \rightarrow \infty$, seemingly 
implying that the equilibrium is reached and therefore $H$ belongs to SGE. 
But, as obvious from $H=H_0+ x V$, 
in limit $x \rightarrow \infty$, $H=x V$ and 
therefore $H$ must be a GG matrix. This contradiction is a result of 
the error made in not ensuring the mean spacing of $H$ same as 
$H_0$ and $V$ \cite{fh}. 
Here, to ensure the latter, we need to use a modified 
Hamiltonian, given by $H= {\rm e}^{-\tau\over N} H_0 + 
{\sqrt{{1-{\rm e}^{-2 \tau\over N}}\over N}} V$ 
with $\tau= - N^{-1} {\rm ln}\;{\rm cos}(x/N)$ (same as before 
in large-$N$ limit). 
The effect of this modification on F-P equation (17) is that now 
${\partial P\over\partial Y } =    
{\partial P\over\partial \tau } + {1\over N (1-{\rm e}^{-2\tau/N})}   
\sum_s \sum_{k \le l} \left(\gamma - y_{kl;s}\right) \left[ 2 y_{kl;s}
{\partial P\over\partial y_{kl;s}}  
- \gamma^{-1} b_{kl;s} {\partial P\over\partial b_{kl;s}} \right]$ 
and the coefficient 
$\gamma$ of the drift term is now replaced by $N^{-1} \gamma$ (see eq.(13) 
of \cite{ps2}). The $Y$ can now be obtained by the second method given in 
section I.C.  
\subsection{Non-Gaussian Ensembles}

As mentioned before, the RM models of complex systems can, in general, 
be non-Gaussian, e.g. $\rho(H)= C {\rm exp}^{-\sum_{k\le l} f(H_{kl})}$ 
with $f$ as an arbitrary function and  it is not an easy task to 
obtain the correlations in this case. However this case can be analyzed 
by our method if $f$ is a well 
behaved function and can be expanded in a Taylor's series.  
To understand this, let us consider an ensemble of real-symmetric matrices 
  $H$ with 
distribution of a more general nature 
e.g. $f$ as a polynomial of $H$ with degree $2M$,  
 $f_{kl}(x)=\sum_{r=1}^M \gamma_{kl}(r)x^{2r}$
with $C$ as the normalization constant
and variances for the diagonal and off-diagonal matrix
elements chosen to be arbitrary.

To obtain an evolution equation in this case, 
we now consider the sum 
 $ 2 \sum_{r=1}^M  r \sum_{k \le l}\left(\gamma - 
y_{kl}(1) \right)y_{kl}(r) 
{\partial \tilde P\over\partial y_{kl}(r)}$
(with $P=C\tilde P$ and $y_{kl}(r)=g_{kl} \gamma_{kl}(r)$) where  
the derivative of $\tilde P$ 
with respect to $\gamma_{kl}(1)$ can be 
shown to be following 
(with $\rho=C\tilde{\rho} $)  

\begin{eqnarray}
{\partial \tilde P\over\partial \gamma_{kl}{(1)}}  
& = &   
 {1\over 2\gamma_{kl}(1)}
\int \prod_{i=1}^N \delta(\mu_i-\lambda_i) 
H_{kl}{\partial\tilde \rho \over \partial H_{kl}} 
{\rm d}H   
-  \sum_{r=2}^M  r { \gamma_{kl}{(r)} \over \gamma_{kl}{(1)}}
\int \prod_{i=1}^N \delta(\mu_i-\lambda_i) 
{\partial \tilde \rho \over \partial \gamma_{kl}{(r)}}
{\rm d}H 
\end{eqnarray}
Now as 
${\partial \tilde \rho\over\partial \gamma_{kl}(r)} = - H_{kl}^{2r} \rho$
and ${\partial \tilde \rho\over\partial H_{kl}} = 
- 2 \sum_{r=1}^M  r \gamma (r) H_{kl}^{2r-1} \rho$, 
the second integral in eq.(51) being  equal to 
 ${\partial \tilde P \over \partial \gamma_{kl}{(r)}}$, 
eq.(11) can be rearranged to show that 
$2 \sum_{r=1}^M r \gamma_{kl}(r) 
{\partial \tilde P\over\partial \gamma_{kl}(r)} = I_{kl} $
  with $I_{kl}$ given by eq.(12),
(but without subscript $(s)$ on quantities). 
The required evolution equation in this case, 
can be obtained from the following equality: 
\begin{eqnarray}  
 2 \sum_{r=1}^M \sum_{k \le l} r \left(\gamma - 
y_{kl}(1)\right)y_{kl}(r) 
{\partial \tilde P\over\partial y_{kl}(r)}
= \gamma \sum_{k\le l} I_{kl} -\sum_{k\le l} y_{kl}(1) I_{kl}
\end{eqnarray}
where, again,  
$\sum_{k\le l} I_{kl}
= \sum_n {\partial \over \partial\mu_n }(\mu_n \tilde P)$ 
and 
\begin{eqnarray}
\sum_{k\le l} y_{kl}(1)  I_{kl}
= -\sum_n {\partial \over \partial\mu_n}
\left[ {\partial \over \partial\mu_n} +
\sum_{m\not=n}{\beta \over {\mu_m-\mu_n}}\right] \tilde P 
+ \sum_{k\le l} J_{kl}   
\end{eqnarray}
with $J_{kl}$ now given by following relation:
\begin{eqnarray}
J_{kl}  
 &=& - \sum_n {\partial \over \partial \mu_n}
\sum_{k\le l} \int \prod_{i=1}^N \delta(\mu_i-\lambda_i) 
{\partial \lambda_n \over \partial H_{kl}} 
\left[\sum_{r=2}^M r y_{kl}(r) H_{kl}^{2r-1}\right]\;\; \rho 
{\rm d}H \\  
&=& g_{kl} 
\sum_{r=1}^M (r+1) y_{kl}(r+1) {\partial \tilde P \over \partial y_{kl}(r)}  
\left[(2r+1) + 2 \sum_{s=1}^M s y_{kl}(s) 
{\partial \tilde P \over \partial y_{kl}(s)}\right]  
\end{eqnarray}

 using these relations as before, 
one again obtains the F-P equation for $\tilde P$ similar to 
eq.(17) with $\beta=1$ and 
${\partial \tilde P\over\partial Y } =   
2\sum_{k \le l} \sum_{r=1}^M h_{kl}(r)
{\partial \tilde P\over\partial y_{kl}(r)}$    
where 
$h_{kl}(r)= 2 r\; y_{kl}(r)\;(\gamma - y_{kl}(1))
+ (r+1)(2r+1)\; y_{kl}(r+1)\; g_{kl}
+ 2 (r+1)\; y_{kl}(r+1)\; g_{kl} \sum_{s=1}^M 
s\; y_{kl}(s)\;{\partial \tilde P\over\partial y_{kl}(r)}$.  
Note the condition for $Y$ here includes terms of type    
${\partial \tilde P\over\partial y_{kl}(r)} 
{\partial \tilde P\over\partial y_{kl}(s)}$ 
and $Y$ can no longer 
be obtained by methods given in section I.C.

\subsection{Block-Diagonal Ensembles}

The eq.(7) and, therefore, 
evolution equation (17) of 
$P(\mu,Y)$ is no longer valid if the matrix $H$ is in a block-diagonal 
form. This is because the eigenvalues belonging to different 
blocks don't repel each other, are not correlated and undergo an evolution 
independent of the other block.   
For this case,  the evolution of 
eigenvalues in each block can be considered separately, leading to one 
F-P equation similar to eq.(17) for each block. 
A detailed discussion of this case in given in \cite{ps1}.

\section{An Alternative Evolution Equation For The Eigenvalues}

In section I.B, the eq.(17)  governing the evolution of the eigenvalues 
was obtained by using the relation (14). However, as obvious from eq.(13),  
$P$ also satisfies the relation 
\begin{eqnarray}
\sum_{k \le l}\left[2 y_{kl;s}
{\partial P\over\partial y_{kl;s}}  
- b_{kl;s} {\partial P\over\partial b_{kl;s}} \right]  
= \sum_{k\le l}  I_{kl;s} 
\end{eqnarray}

and, therefore, one can define a function $Z(y_{kl;s},b_{kl;s})$ 
such that

\begin{eqnarray}
{\partial P\over\partial Z}  
=\sum_{n}{\partial \over \partial\mu_n}\left(\mu_n P\right)
\end{eqnarray}

Here $Z$ is given by the condition 
${\partial P\over\partial Z}=   
\sum_{k \le l}\left[2 y_{kl;s}
{\partial P\over\partial y_{kl;s}}  
- b_{kl;s} {\partial P\over\partial b_{kl;s}} \right] $ which 
can be solved (as in section II) to show that 
$Z={1\over 4 N^2} 
{\rm ln}\left[\prod_{k\le l} \prod_{s=1}^2 |y_{kl;s}| b_{kl;s}^{-2}\right] 
+ C$. 

	The eq.(57) also describes the evolution of eigenvalues for the same 
ensemble (3). But now the "time"-scale is such that the  
eigenvalues seem to be drifting only, hiding the repulsion between them. 
Again the steady state of eq.(57) is given by $|Z-Z_0|\rightarrow  \infty$ 
and the final 
ensemble as Poisson (with finite, non-zero variances 
for diagonal 
matrix elements and zero variances for the off-diagonal ones). The 
ensemble G will now appear as an intermediate point in a transition from 
 some initial ensemble 
$\rightarrow$ Poisson ensemble and, in principle, the transition can be 
used for the 
analysis of $G$. 
For example, the critical parameter for Anderson transition 
(same model as used in section III) 
can be obtained by taking the initial state "O" as metal with energy level 
distribution described by a 
GUE ($<O_{ii}^2>=\alpha_M^{-1}$, $<O_{ij}^2>=0$) and all $<O_{ij;s}>=t_M$ 
which gives
$Z_0={1\over 2 N^2}\left[N 
{\rm ln} \alpha_M -2 K {\rm ln}| t_M|\right] + C$
The critical region will therefore occur as as intermediate point in 
the GUE $\rightarrow$ Poisson transition with transition parameter 
$\Lambda =D^{-2} (Z-Z_0) =
{1\over 2 N}\left[ N {\rm ln} 
{\alpha \over \alpha_M} - 2 K {\rm ln}{|t|\over | t_M|}\right]$. 
 As obvious, the 
increase of diagonal disorder (${\alpha\over \alpha_M} < 1$) for a fixed 
hopping rate ($t=t_M$) will ultimately lead to Poisson statistics, implying 
localization of states; note here the transition occurs backwards in "time" 
$\Lambda$.  
 However the results for correlations associated with 
SGE $\rightarrow$ Poisson transition are not known which leaves eq.(17) as 
a better tool to analyze the properties of GGEs. . 
The eq.(17) has one more advantage over eq.(58): the reduction of former 
to CSM 
Hamiltonian reveals the underlying 
universality of statistical formulation among various complex system.

\section{Conclusion}

	In this paper, we have described a new method to analyze the 
statistical properties of the RM model of complex systems. Our technique 
is based on the exact reduction of spectral analysis in the general 
case to the one in SGE. This greatly reduces the degree of difficulty of 
the original problem as many of the properties of SGE are already known. 
This also indicates that a thorough knowledge of the properties of SGE 
or CSM will be highly advantageous even for systems with interactions too 
intricate to be modeled by SGE. So far, the probing of GGE is carried out 
only by SUSY technique which requires a saddle point approximation at 
various steps and is not easily applicable, even approximately, to 
cases where our technique can be used for exact probing. Note the main 
term in GGEs responsible for the correspondence with CSM Hamiltonian  
is due to the repulsion between eigenvalues. As the mathematical origin of 
this term lies in the transformation from matrix space to eigenvalue space 
which is same for all the hermitian ensembles (belonging to same symmetry 
class), the correspondence with 
CSM Hamiltonian should exist for almost all of them irrespective of the 
distribution of their matrix elements. As discussed in section III, 
our study also confirms the conjecture 
regarding the one parameter scaling of localization and provides the 
formula for relevant parameter.

         The reduction technique presented here raises some basic 
questions. Why the reparametrization of the spectral properties of different 
RM ensembles
results in to a similar  mathematical formulation for them? 
In other words, why the eigenvalues of quantum operators associated with 
complex systems evolve in a similar ordered way  
(like equations of motion for Calogero particles) notwithstanding the varied 
nature of their complexity? The reason may lie in the following. 
The eigenvalues and eigenfunctions of a Hamiltonian evolve due to a change 
in either degree or nature of its complexity. The evolution of the 
eigenfunction is chaotic in the sense that the overlapping between the 
eigenfunctions, associated with two Hamiltonians even with slightly different 
complexity, decreases rapidly in time (page 2 of \cite{fh}). However 
an eigenvalue 
of an operator is its average value in the state described by the associated 
eigenfunction and an ordered evolution of the former will, in 
general, imply an 
ordered change in the average behavior of the latter. 
Thus it seems that the eigenvalues 
and eigenfunctions, on an average, are not able to view the fine 
subtlities of the varied nature of 
complexity  and therefore are not affected too drastically to loose 
correlations even when nature of the complexity changes. Note, for a small 
change in the interactions, this result is not surprising and used as the 
base for the perturbation theory. 
But the results in this paper imply that the eigenvalues (and physics based on 
them) even after a violent change in  the interactions remain correlated 
in the parametric space. 
Thus it seems that certain physical properties, based on average behavior of 
eigenvalues and eigenfunctions,  of one complex system are related 
to the physics of another, very different in nature of the interactions.

I am grateful to B.S.Shastry and N.Kumar    
for various useful suggestions during the course of this study. I would 
also like to thank K.Frahm and V. Kravatsov for useful criticism of the work.  
A brief 
discussion with B. Altschuler and B. Huckenstein has also been helpful.

\begin{appendix}

\section{\bf Proof of Eqs.(3,4,5)}  
\vspace{.5in}

The use of the eigenvalue equation $H U= U \Lambda $, with $U$ as a 
unitary matrix and $\Lambda$ the eigenvalue matrix, leads to following: 
     
\begin{eqnarray}
 \sum_j H_{ij} U_{jn} = \lambda_n U_{in} 
 \quad {\rm and} \quad \sum_i H_{ij} U_{in}^* = \lambda_n U_{jn}^* 
\end{eqnarray}

where $H_{ij}=H_{ij;1} + i H_{ij;2}$. Differentiating both sides of above 
equation with respect to $H_{kl;s}$ 
(with $s=1\; {\rm or}\; 2$), we get

\begin{eqnarray}
\sum_j {\partial U_{jn}\over\partial H_{kl;s}} H_{ij} + 
\sum_j U_{jn} {\partial H_{ij}\over\partial H_{kl;s}}  &=& 
\lambda_n {\partial U_{in}\over\partial H_{kl;s}}  +  
{\partial \lambda_n \over\partial H_{kl;s}} U_{in}  
\end{eqnarray}

Now as $\sum_i U_{in}^* U_{im} = \delta_{nm}$, 
multiplying both the sides by $U_{in}^{*}$  followed by a 
summation 
over all $i$'s, we get the following

\begin{eqnarray}
 {\partial \lambda_n \over\partial H_{kl;s}} &=& 
\sum_{i,j} U_{in}^{*} {\partial H_{ij}\over\partial H_{kl;s}} U_{jn}
\end{eqnarray}

which further gives 

\begin{eqnarray}
 {\partial \lambda_n \over\partial H_{kl;s}} &=& 
 i^{s-1}{1\over g_{kl}}
 \left[ U_{ln} U_{kn}^{*} - (-1)^s U_{ln}^{*} U_{kn}\right]
\end{eqnarray}

This can further be used to show that

\begin{eqnarray}
\sum_{k\le l}
 \sum_{s=1}^2 {\partial \lambda_n \over\partial H_{kl;s}} H_{kl;s}  
&=& \sum_{k\le l}  
 {1\over g_{kl}}
\left[  U_{ln} U_{kn}^{*} \sum_s i^{s-1} H_{kl;s} +  
U_{ln}^{*} U_{kn} \sum_s i^{s-1} (-1)^{s+1} H_{kl;s}\right]\\
&=& \sum_{k\le l}  
 {1\over g_{kl}}
\left[ H_{kl} U_{ln} U_{kn}^{*} + H_{kl}^* U_{ln}^{*} U_{kn}\right]\\
&=& \sum_{k\le l}  
 {1\over g_{kl}}
 H_{kl} U_{ln} U_{kn}^{*} 
 +  \sum_{k\ge l}  
 {1\over g_{kl}}
 H_{lk}^* U_{kn}^{*} U_{ln}\\
&=& \sum_{k, l}  
 H_{kl} U_{ln} U_{kn}^{*} = \lambda_n 
\end{eqnarray}
where eq.(A8) is obtained from eq.(A7) by using Hermitian properties of 
$H$ ($H_{lk}^* = H_{kl}$). 
By using eq.(A4), One can also show that 

\begin{eqnarray}
\sum_{k\le l} g_{kl}
 \sum_{s=1}^2 {\partial \lambda_n \over\partial H_{kl;s}}   
  {\partial \lambda_m \over\partial H_{kl;s}}   
&=& \sum_{k\le l} \sum_{s=1}^2 
 i^{2(s-1)}{1\over g_{kl}}
\left[ U_{ln} U_{kn}^{*} - (-1)^s U_{ln}^{*} U_{kn}\right]
\left[ U_{lm} U_{km}^{*} - (-1)^s U_{lm}^{*} U_{km}\right]\\
&=& \sum_{k\le l}  
 {2\over g_{kl}}
\left[ U_{ln} U_{kn}^{*} U_{km} U_{lm}^* + 
 U_{ln}^{*} U_{kn} U_{lm} U_{km}^* \right]\\
&=& 2 \sum_{k, l}  
 U_{ln} U_{kn}^{*} U_{km} U_{lm}^*  =
\sum_{k}  U_{kn} U_{km}^{*} \sum_{l} U_{lm} U_{ln}^* = 2\delta_{mn}
\end{eqnarray}

where eq.(A11) follows from eq.(A10) by writing
 $ \sum_{k\le l}  
 U_{ln}^* U_{kn} U_{km}^* U_{lm}  =
  \sum_{k\ge l}  
 U_{ln} U_{kn}^{*} U_{km} U_{lm}^* $ and the last equality in eq.(A11) is 
due to unitary nature of $U$. 

\section{\bf Proof of Eq.(6)}

Multiplying both the sides of eq.(A2) by $U_{im}^{*}$ ($m \not=n$) 
followed by a 
summation 
over all $i$'s, we get the following

\begin{eqnarray}
\sum_j U_{jm}^{*} {\partial U_{jn}\over\partial H_{kl;s}} &=& 
{1\over {\lambda_n -\lambda_m}}
\sum_{i,j} U_{im}^{*} {\partial
 H_{ij}\over\partial H_{kl;s}} U_{jn}
\end{eqnarray}

a multiplication of both the sides by $U_{rm}$ followed by a summation 
over all $m$'s then gives

\begin{eqnarray}
 {\partial U_{rn}\over\partial H_{kl;s}} &=& i^{s-1} {1\over g_{kl}} 
\sum_{m\not=n} {U_{rm}\over {\lambda_n -\lambda_m}}
 \left( U_{km}^{*} U_{ln} - (-1)^s U_{lm}^{*} U_{kn} \right)
\end{eqnarray}

\section{\bf Proof of Eq.(7)}

\begin{eqnarray}
\sum_{k\le l} g_{kl} 
 \sum_{s=1}^2 {\partial^2 \lambda_n \over\partial H_{kl;s}^2}   
&=& \sum_{k\le l} \sum_{s=1}^2 
 i^{s-1}{1\over g_{kl}}
  {\partial  \over\partial H_{kl;s}}   
\left[ U_{ln} U_{kn}^{*} - (-1)^s U_{ln}^{*} U_{kn}\right]\\
&=& \sum_{k\le l} \sum_{s=1}^2 i^{s-1} 
  \left[{\partial U_{kn}^* \over\partial H_{kl;s}} U_{ln}  
 + {\partial U_{ln} \over\partial H_{kl;s}} U_{kn}^*  
 + (-1)^{s+1} {\partial U_{ln}^* \over\partial H_{kl;s}}   U_{kn}
 + (-1)^{s+1}{\partial U_{kn} \over\partial H_{kl;s}} U_{ln}^* \right] 
\end{eqnarray}

Now by using eq.(B2) and its complex conjugate in eq.(C2) and by summing over 
$s$, we get

\begin{eqnarray}
\sum_{k\le l} g_{kl} 
 \sum_{s=1}^2 {\partial^2 \lambda_n \over\partial H_{kl;s}^2}   
&=& 4 \sum_{k\le l} {1\over g_{kl}} \sum_m {1\over \lambda_n -\lambda_m} 
\left[ U_{km} U_{km}^* U_{ln} U_{ln}^* 
+ U_{kn} U_{kn}^* U_{lm} U_{lm}^* \right]\\  
&=& 4 \sum_{k, l}  \sum_m {1\over \lambda_n -\lambda_m} 
\left[ U_{km} U_{km}^* U_{ln} U_{ln}^* \right]\\ 
&=& 4 \sum_m {1\over \lambda_n -\lambda_m} 
\left[\sum_k U_{km} U_{km}^*\right] \left[\sum_l U_{ln} U_{ln}^* \right] 
\end{eqnarray}

Now by using the unitary relation $\sum_j U_{jm}^* U_{jm} = 1$, 
one  obtains the desired relation (7).

..

\section{\bf Proof of Eq.(15)  }

The  eq.(12) gives us the following,

\begin{eqnarray}
\sum_{k\le l} \sum_{s=1}^2 I_{kl;s}
&=&\sum_n {\partial \over \partial\mu_n}
\int \prod_{i}\delta(\mu_i-\lambda_i)
\left[ \sum_{k\le l}
\sum_{s=1}^2 {\partial \lambda_n\over \partial H_{kl;s}}
 H_{kl;s}\right] 
\rho {\rm d}H\
\end{eqnarray}
The use of eq.(A8) will further simplify it in following form 

\begin{eqnarray}
\sum_{k\le l} \sum_{s=1}^2 I_{kl;s}
&=&\sum_n {\partial \over \partial\mu_n}
\int \prod_{i}\delta(\mu_i-\lambda_i)
\lambda_n \rho {\rm d}H\\
&=&\sum_n {\partial \over \partial\mu_n}(\mu_n P)
\end{eqnarray}

\section{\bf Proof of eq.(16)  }

For each $s$-value, we have the following relation

\begin{eqnarray}
\sum_{k\le l}  y_{kl;s}  I_{kl;s}
&=&\sum_{n=1}^N {\partial \over \partial \mu_n} 
\sum_{k\le l} g_{kl} \alpha_{kl;s} 
\int \prod_{i=1}^N 
\delta(\mu_i-\lambda_i) 
{\partial \lambda_n \over \partial H_{kl;s}} 
 H_{kl;s} \rho {\rm d}H\\
&=& -\sum_{n=1}^N {\partial \over \partial \mu_n} 
\sum_{k\le l} {g_{kl} \over 2}
\int \prod_{i=1}^N 
\delta(\mu_i-\lambda_i)  
{\partial \lambda_n \over \partial H_{kl;s}}  
\left[{\partial \over \partial H_{kl;s}} - 
2\alpha_{kl;s} b_{kl;s} \right] \rho 
 {\rm d}H \\
&=& -\sum_{n=1}^N {\partial \over \partial \mu_n} 
\sum_{k\le l}  {g_{kl}\over 2}
\int \prod_{i=1}^N 
\delta(\mu_i-\lambda_i)  
{\partial \lambda_n \over \partial H_{kl;s}}  
{\partial \rho \over \partial H_{kl;s}} 
 {\rm d}H 
+ \sum_{k \le l} J_{kl;s}
\end{eqnarray} 

where eq.(E3) is obtained by using the equality:
${\partial \rho\over \partial H_{kl;s}}= 
-2\alpha_{kl;s}(H_{kl;s} - b_{kl;s}) \rho$ and 
$J_{kl;s}$ is given by eq.(E9).

 By integrating eq.(E3) further 
by parts, one obtains

\begin{eqnarray}
\sum_{k\le l} \sum_{s=1}^2 y_{kl;s}  I_{kl;s}
&=& \sum_{s} \sum_n {\partial \over \partial\mu_n}
\sum_{k\le l}  {g_{kl}\over 2}
\int \left({\partial\over\partial H_{kl;s}}
\prod_{i}\delta(\mu_i-\lambda_i)\right)
{\partial \lambda_n \over \partial H_{kl;s}}  
\rho {\rm d}H\\
&+& \sum_s \sum_n {\partial \over \partial\mu_n}
\sum_{k\le l}{g_{kl}\over 2} \int \prod_{i}\delta(\mu_i-\lambda_i)
 {\partial^2 \lambda_n\over\partial H_{kl;s}^2}
 \rho {\rm d}H
+ \sum_{k \le l} \sum_s J_{kl;s}\\
&=&-\sum_n {\partial \over \partial\mu_n}
\sum_m {\partial \over \partial\mu_m}
\int\prod_{i}\delta(\mu_i-\lambda_i)
\left[\sum_s \sum_{k\le l} {g_{kl}\over 2} 
{\partial \lambda_m \over \partial H_{kl;s}}
{\partial \lambda_n \over \partial H_{kl;s}}  
\right] \rho {\rm d}H\\
&-&\sum_n {\partial \over \partial\mu_n}
\int \prod_{i}\delta(\mu_i-\lambda_i)
\left[\sum_{m \not= n}{2\over {\lambda_m -\lambda_n}}\right]
 \rho(H) {\rm d}H
+  \sum_{k \le l} \sum_s J_{kl;s}\\
&=&-\sum_n {\partial^2 P \over \partial\mu_n^2}
-\sum_n {\partial \over \partial\mu_n}
\left[2 \sum_{m \not= n}{P\over {\mu_m -\mu_n}}\right]
+ \sum_{k \le l}\sum_s J_{kl;s}
\end{eqnarray}


where $J_{kl;s}$ can be obtained as follows:

\begin{eqnarray} 
 J_{kl;s}&=&  
 y_{kl;s} b_{kl;s}    
\sum_{n=1}^N {\partial \over \partial \mu_n} 
\int \prod_{i=1}^N 
\delta(\mu_i-\lambda_i) 
{\partial \lambda_n \over \partial H_{kl;s}}  
\rho {\rm d}H\\
&=&- y_{kl;s} b_{kl;s}    
\int  
{\partial \prod_{i=1}^N \delta(\mu_i-\lambda_i) \over \partial H_{kl;s}}  
\rho {\rm d}H\\
&=& y_{kl;s} b_{kl;s}    
\int \prod_{i=1}^N 
\delta(\mu_i-\lambda_i)   
{\partial \rho \over \partial H_{kl;s}}  
 {\rm d}H\\
&=&- y_{kl;s} b_{kl;s}    
\int \prod_{i=1}^N 
\delta(\mu_i-\lambda_i)   
{\partial \rho \over \partial b_{kl;s}}  
 {\rm d}H 
=- y_{kl;s} b_{kl;s}    
{\partial P \over \partial b_{kl;s}}  
\end{eqnarray}

where in eq.(A32), the equality 
${\partial \rho \over \partial b_{kl;s}}= 
2\alpha_{kl;s} (H_{kl;s} - b_{kl;s}) \rho  
= -{\partial \rho \over \partial H_{kl;s}}$ is used.  
A substitution of eq.(E12) in eq.(E8) now leads to the eq.(16). 

..

\section{\bf A General Method to Obtain $Y$}

Let us consider a transformation of 
$M={2 N^2}$ 
coordinates $\{r_j\}$ to 
another set of $M$ coordinates $\{Y_i\}$, where $r_j$'s are various 
coefficients $y_{kl;s}$ (total $N^2$) and $b_{kl;s}$ 
(total $ N^2$).  
The $Y_i$'s should be chosen such that the 
 right hand side of the eq.(18), summing over all  
$y_{kl;s}$,'s and $b_{kl;s}$'s can be rewritten as 

\begin{eqnarray}
\sum_{i}^{M} {\partial P\over\partial Y_i} 
= \sum_{k \le l}  2\left(\gamma - y_{kl;s}\right) y_{kl;s} 
{\partial P\over\partial y_{kl;s}} 
- \gamma  \sum_{k\le l}  b_{kl;s} 
{\partial P\over\partial b_{kl;s}} \equiv   
\sum_{j=1}^{M} g_j (r_1,r_2,..,r_M) {\partial P\over\partial r_j} 
\end{eqnarray}
where, for our case,
$g_i(r_1,..,r_M) =   2\left(\gamma - r_i\right) r_i$ 
 if $r_i$ is one of the $y_{kl;s} $,
 and, $g_i(r_1,..,r_M) = -\gamma r_i $  
   if $r_i$ is one of the $b_{kl;s}$. 

Now, as we want  
$\sum_{i}^{M} {\partial \over\partial Y_i} 
= {\partial \over\partial Y_1}$, with $Y_1 \equiv Y$, this imposes 
 following conditions on the functions $Y_i$'s 
(as can be shown by using  the theory of partial differentiation)  
 
\begin{eqnarray}
 {\partial P\over\partial Y_1}   
=\sum_{i=1}^M \sum_{j}^M g_j (r_1,r_2,..,r_M) {\partial P\over\partial Y_i} 
{\partial Y_i\over\partial r_j}    
\end{eqnarray}
and therefore 
\begin{eqnarray}
\sum_{j=1}^M g_j(r_1,r_2,..,r_M)
 {\partial Y_i \over \partial r_j} = \delta_{1i}  
\end{eqnarray}

According to theory of partial differential equations 
\cite{sne},  
the general solution of linear PDE 
$\sum_{i}^M P_i(x_1,x_2,..,x_M){\partial Z \over \partial x_i} = R  $
is $F(u_1,u_2,..,u_n)=0$ where $F$ is an arbitrary function and 
$u_i(x_1,x_2,..,x_n,Z)=c_i$ (a constant), $i=1,2,..,n$ are independent 
solutions of the following equation 

\begin{eqnarray}
{{\rm d}x_1 \over P_1} = 
{{\rm d}x_2 \over P_2} =..... 
{{\rm d}x_k \over P_k} =......
{{\rm d}x_M \over P_M} = 
{{\rm d}Z \over R}  
\end{eqnarray}

Thus the general 
solution of eq.(F3) for each $Y_j$ is given by a relation 
$F_j(u_{1j},u_{2j},..,u_{Mj}) = 0$ 
where  function $F_j$ is arbitrary and $u_{ij}(r_1, r_2, ..,r_M,Y_j)=c_{ij}$, 
$(i=1,2,..,M)$ (with $c_{ij}$'s as constants) are independent solution of the 
equation 

\begin{eqnarray}
{{\rm d}r_1 \over g_1} = 
{{\rm d}r_2 \over g_2} =..... 
{{\rm d}r_k \over g_k} =......
{{\rm d}r_M \over g_M} = 
{{\rm d}Y_j \over \delta_{1j}}  
\end{eqnarray}

The above set of equations can be solved  for various $Y_j$ to 
obtain $F_j$.   
For $Y_1$, we get the  relations 
$Y_1 - {1\over 2}{\rm log}{r_i\over |r_i -\gamma|}= 
c_{i1}\quad ({i=1,..,M/2})$,  
$Y_1 + {1\over \gamma}{\rm log}|{r_i}|= c_{i1}\quad ({i=1+M/2,..,M})$,  
and therefore $F_1$ satisfies the relation
$F_1 (
Y_1 - {1\over 2}{\rm log}{r_1\over |r_1 -\gamma|},..,   
Y_1 - {1\over 2}{\rm log}{r_{M/2}\over |r_{M/2} -\gamma|}),  
Y_1 + \gamma^{-1} {\rm log}|{r_{M/2}|},..,   
Y_1 + \gamma^{-1} {\rm log}|{r_M}|)=0$. 
The function $F_1$ being arbitrary here, 
this relation can also be expressed in the follwing form:
   
\begin{eqnarray}
Y_1 ={1\over M}\left[{1\over 2}
\sum_{i=1}^{ M/2} {\rm log}{r_i\over |r_i -\gamma|} - 
{1\over \gamma} \sum_{i=M/2+1}^M {\rm log}\; |r_i|\right]   + C
\end{eqnarray}
where $C$ is another arbitrary function of constants: for example  
$C \equiv C ({1\over 2}
 {\rm log}{r_1\over |r_1 -\gamma|}   
 + \gamma^{-1} {\rm log}|r_M|, 
 {1\over 2}{\rm log}{r_2\over |r_2 -\gamma|}   
 + \gamma^{-1} {\rm log}|r_M|,...,
 {1\over 2}{\rm log}{r_{M-1}\over |r_{M-1} -\gamma|}   
 + \gamma^{-1} {\rm log}|r_M |)$. 

Similarly the
 variables $Y_i$, $i>1$, can be obtained however their knowledge is not 
 required for our analysis. 

\section{ }

	The choice of $\gamma$ is based only on the requirement that 
$y_{kl}(O) > y_{kl}(G) > \gamma$ for all $k,l$. 
Thus $\gamma$ can take any value 
such that $\gamma \le {\rm min}y_{kl}(G)$. Let us consider two such 
possibilities  for $\gamma$, $\gamma=\gamma_1$ and $\gamma=\gamma_2$
and try to evaluate properties of $G$  on these curves referred as $T1$ 
and $T2$ respectively. Let the value of 
$Y$ for $G$ on these curves be $Y_1$ and $Y_2$ where 
 
\begin{eqnarray}
Y_1 &=& {1\over 2 N^2} \sum_{k\le l} \sum_{s=1}^2 \left[ {1\over 2}
 {\rm ln}{y_{kl;s}\over |y_{kl;s}-\gamma_1|} - {1\over \gamma_1}
{\rm ln} b_{kl;s}\right] +C\\ 
Y_2 &=& {1\over 2 N^2} \sum_{k\le l} \sum_{s=1}^2 \left[ {1\over 2}
 {\rm ln}{y_{kl;s}\over |y_{kl;s}-\gamma_2|} - {1\over \gamma_2}
{\rm ln} b_{kl;s}\right] +C 
\end{eqnarray}

However $Y_1$ can also be written as follows

\begin{eqnarray}
Y_1 &=& {1\over 2 N^2} \sum_{k\le l} \sum_{s=1}^2 \left[ {1\over 2}
 {\rm ln}{y_{kl;s}'\over |y_{kl;s}'-\gamma_2|} - {1\over \gamma_2}
{\rm ln} b_{kl;s}'\right] +C 
\end{eqnarray}

Now as $y_{kl;s}'= y_{kl;s} {\gamma_2\over \gamma_1} \not= y_{kl;s}$, this 
implies that $Y_1$ would correspond to a point, different from $Y_2$, on 
the transition 
curve $T_2$ and therefore would give  
properties for the ensemble $G$ different from those given by $Y_2$. 
This conclusion is, however, erroneous 
and is a result of the rescaling applied only to one point $Y_1$ on the 
transition curve $T_1$.   To get the right answer, the whole 
curve $T_1$ 
should be rescaled 
which would require a rescaling of the end-points too and 
therefore changed distances on the rescaled curve (call it $T_1'$). Thus 
the point $Y_1$ will appear at the same location on $T_1'$-curve, 
relative to end-points, where $Y_2$ appears on $T_2$-curve  and therefore 
both will imply the same properties for the ensemble $G$.


\end{appendix}


\begin{references}
\bibitem{cal}
F.Calogero, J. Math. Phys., 10, 2191, 2197 (1969).
\bibitem{gmw}
T.Guhr, G.A. Muller-Groeling and H.A. Weidenmuller, 
Phys. Rep. V299, 189, (1998).
\bibitem{sla}
B.D.Simon, P.A.Lee, B.L.Altshuler, Phys.Rev.Lett, 72, 64,(1994).
\bibitem{ef}
K.B.Efetov, Adv. Phys. 32, 53, (1983).
\bibitem{dy}
F.Dyson, J. Math. Phys. 3, 1191 (1962).
\bibitem{fh}
F.Haake, {\bf Quantum Signature of Chaos}, Springer, Berlin (1991).
\bibitem{os}
O.Narayan and B.S.Shastry, Phys. Rev. Lett., 71, 2106, (1993). 
\bibitem{bs}
B.S.Shastry, Proceedings of $16^{\rm th}$ Taniguchi International Symposium 
on the theory of Condensed Matter: "Correlated effects in low dimensional 
systems", October 23-29, 1993, Shima, Japan, 
(Springer Verlag 1994, Eds: N.Kawakami and A.Okiji). 
\bibitem{sut}
B.Sutherland, J. Math. Phys., 12, 246, (1971); 12, 252, (1971); 
Phys. Rev. A, 4, 2019, (1971); 5, 1372, (1972).
\bibitem{las}
S.M.Bhattacharjee and S.Mukherjee, Phys. Rev. Lett. 83, (1999);
S.M.Bhattacharjee, Phys. Rev. Lett. 76, (1996);
M.Lassig, Phys. Rev. Lett., 77, 526, (1996).
\bibitem{pec}
P.Pechukas, Phys. rev. Lett., 51, 943, (1983); 
S.Wojciechowski, Phys. Lett., 111A, 101, 1985; 
T.Yukawa, Phys. Rev. Lett., 54, 1883, (1985); 
K.Nakamura and M.Lakshmanan, Phys. Rev. Lett., 57, 1661, (1986).
\bibitem{meta}
M.L.Mehta, Random Matrices (Acedemic Press, Boston) (1991).
\bibitem{ap}
A.Pandey, Chaos, Soliton and Fractals, 5, (1995). 
A.Pandey and P.Shukla, J. Phys. A, (1991).
\bibitem{ps1}
P.Shukla, to appear in Physica E, 2000 and Physica A, 2000.
\bibitem{sne}
Snedden,{\bf Elements of Partial Differential Equations}, McGraw-Hill, (1988).
\bibitem{itz}
C.Itzykson and J.B.Zuber, J. Math. Phys. 21, 411 (1980).
\bibitem{fkpt}
J.B.French, V.K.B.Kota, A.Pandey and S.Tomsovic, 
Ann. Phys. (N.Y.), 181, (1988).
\bibitem{klh}
M.Kus, M.Lewenstein and F.Haake, Phys. rev. A, 44, 2800, (1994). 
\bibitem{znc}
Z.N.C. Ha, Phys. rev. Lett. 73, 1574 (1994)  and 
Nucl. Phys. B 435 604 (1995). ; 
\bibitem{and}
P.W.Anderson, Phys. Rev. 19, 1492, (1958).
\bibitem{izra}
F.M.Izrailev, Phys. rep. 196, 299 (1990).
\bibitem{fm2}
Y.V.Fyodorov and A.D. Mirlin,  Int. J. Mod. Phys. B, 8, 3795 (1994).   
\bibitem{fm91}
Y.V.Fyodorov and A.D. Mirlin, Phys. Rev. Lett., 67, 2405 (1991).   
\bibitem{pr}
T.Prosen and M.Robnik, J.Phys.A 26, 1105, (1993).
\bibitem{ajs}
A.Altland, M.Janssen and B.Shapiro, Phys. Rev. E 56, 1471 (1997). 
\bibitem{mfdqs}
A.D.Mirlin, Y.V.Fyodorov, F-M.Dittes, J.Quezada and T.H.Seligman,  
Phys. Rev. E 54 3221 (1996).
\bibitem{fggk}
V.V.Flambaum, A.A.Gribakina, G.F.Gribakin and M.G.Kozolov, 
 Phys. Rev. A 50 267 (1994).
\bibitem{wig}
E.Wigner, Ann. Math. 62, 548 (1955); 65, 203 (1957).
\bibitem{fcic}
Y.V.Fyodorov, O.A.Chubykalo, F.M.Izrailev and G.Casati, 
Phys. Rev. Lett. 76 1603 (1996).
\bibitem{shep}
D.L.Sheplyansky, Phys. Rev. Lett. 73, 2607 (1994).
\bibitem{bh}
B.Huckestein, Rev. Mod. Phys. 67, 1995. 
\bibitem{bh1}
 B.Huckestein and L.Schweitzer, Phys. Rev. Lett. 72, 713, (1994). 
\bibitem{klaa}
V.E.Kravtsov, I.V.Lerner, B.L.Altshuler and A.G.Aronov, 
Phys. Rev. Lett. 72, 888 (1994)
\bibitem{km}
V.E.Kravtsov and K.A.Muttalib, Phys. Rev. Lett. 79, 1913, (1997).
\bibitem{ckl}
J.T.Chalker, V.E.Kravtsov and I.V.Lerner, Pisma Zh. Eksp. Teor. Fiz. 
64, 355 (1996).
\bibitem{bog}
E.B.Bogomolny, U.Gerland and C.Schmit, Phys. Rev. E59, R1315 (1999). 
\bibitem{mcin}
K.A.Muttalib, Y.Chen, M.E.H.Ismail and V.N.Nicopoulos, 
Phys. Rev. Lett. 71, 471 (1993).
\bibitem{mns}
M.Moshe, H.Neuberger and B.Shapiro, Phys. Rev. Lett. 73, 1497 (1994).
\bibitem{fgm}
K.M.Frahm, T.Guhr, A.Muller-Groeling,  Ann. Phys. (N.Y.) 270, 292 (1998).  
\bibitem{ks}
H.Kunz and B.Shapiro, Phys. Rev. E, 58, 400, 1998.
\bibitem{mir}
F.Evers and A.D.Mirlin, Cond-Mat/0001086
\bibitem{ps2}
P.Shukla, Phys. Rev. E 59, 5207, 1999.  
\bibitem{fcic}
Y.V.Fyodorov, O.A.Chubykalo, F.M.Izrailev and G.Casati, 
Phys. Rev. Lett. 76 1603 (1996).
\bibitem{}
* E-Mail : Shukla@phy.iitkgp.ernet.in
\end{references}
\end{document}